\documentclass[namedreferences]{solarphysics}
\usepackage[hyperref,optionalrh,solaromanenum]{spr-sola-addons} 
\usepackage{graphicx}                    
\usepackage{hyperref}
\hypersetup{
    colorlinks=true,
    linkcolor=blue,
    citecolor=blue,
    filecolor=blue,
    urlcolor=blue
}

\begin{document}

\begin{article}

\begin{opening}

\title{CMEs in the Heliosphere: III. A Statistical Analysis of the Kinematic Properties Derived from Stereoscopic Geometrical Modelling Techniques Applied to CMEs Detected in the Heliosphere from 2008 to 2014 by STEREO/HI-1}


\author[addressref={aff1},corref,email={david.barnes@stfc.ac.uk}]{\fnm{D.}~\lnm{Barnes}\orcid{0000-0003-1137-8220}}
\author[addressref={aff1}]{\fnm{J.A.}~\lnm{Davies}\orcid{0000-0001-9865-9281}}
\author[addressref={aff1}]{\fnm{R.A.}~\lnm{Harrison}\orcid{0000-0002-0843-0845}}
\author[addressref={aff1}]{\fnm{J.P.}~\lnm{Byrne}}
\author[addressref={aff1}]{\fnm{C.H.}~\lnm{Perry}\orcid{0000-0001-5083-6808}}
\author[addressref={aff2}]{\fnm{V.}~\lnm{Bothmer}\orcid{0000-0001-7489-5968}}
\author[addressref={aff3}]{\fnm{J.P.}~\lnm{Eastwood}\orcid{0000-0003-4733-8319}}
\author[addressref={aff4,aff5}]{\fnm{P.T.}~\lnm{Gallagher}\orcid{0000-0001-9745-0400}}
\author[addressref={aff6}]{\fnm{E.K.J.}~\lnm{Kilpua}\orcid{0000-0002-4489-8073}}
\author[addressref={aff7}]{\fnm{C.}~\lnm{M\"ostl}\orcid{0000-0001-6868-4152}}
\author[addressref={aff8}]{\fnm{L.}~\lnm{Rodriguez}\orcid{0000-0002-6097-374X}}
\author[addressref={aff9}]{\fnm{A.P.}~\lnm{Rouillard}}
\author[addressref={aff10}]{\fnm{D.}~\lnm{Odstr\v{c}il}\orcid{0000-0001-5114-9911}}

\address[id=aff1]{STFC RAL Space, Rutherford Appleton Laboratory, Harwell Campus, Oxfordshire, OX11 0QX, UK}
\address[id=aff2]{Institue for Astrophysics, University of G\"ottingen, 37077 G\"ottingen, Germany}
\address[id=aff3]{Blackett Laboratory, Imperial College London, SW7 2AZ, UK}
\address[id=aff4]{School of Physics, Trinity College Dublin, Dublin 2, Ireland}
\address[id=aff5]{School of Cosmic Physics, Dublin Institute for Advanced Studies, Dublin 2, Ireland}
\address[id=aff6]{Department of Physics, University of Helsinki, PO Box 64, 00014, Helsinki, Finland}
\address[id=aff7]{Space Research Institute, Austrian Academy of Sciences, Schmiedlstrasse 6, 8042 Graz, Austria}
\address[id=aff8]{Royal Observatory of Belgium, Ringlaan 3, 1180 Brussels, Belgium}
\address[id=aff9]{Institut de Recherche en Astrophysique et Planetologie, 9 Ave du Colonel Roche, 31028 Toulouse Cedex 4, France}
\address[id=aff10]{School of Physics, Astronomy and Computational Sciences, George Mason University, Fairfax, VA 22030-4444, USA}

\runningauthor{D. Barnes \emph{et al.}}
\runningtitle{CMEs in the Heliosphere: III}

\begin{abstract}
We present an analysis of coronal mass ejections (CMEs) observed by the \emph{Heliospheric Imagers} (HIs) on board NASA's \emph{Solar Terrestrial Relations Observatory} (STEREO) spacecraft. Between August 2008 and April 2014 we identify 273 CMEs that are observed simultaneously, by the HIs on both spacecraft. For each CME, we track the observed leading edge, as a function of time, from both vantage points, and apply the Stereoscopic Self-Similar Expansion (SSSE) technique to infer their propagation throughout the inner heliosphere. 
The technique is unable to accurately locate CMEs when their observed leading edge passes between the spacecraft, however, we are able to successfully apply the technique to 151, most of which occur once the spacecraft separation angle exceeds 180$^\circ$, during solar maximum. We find that using a small half-width to fit the CME can result in observed acceleration to unphysically high velocities and that using a larger half-width can fail to accurately locate the CMEs close to the Sun because the method does not account for CME over-expansion in this region. Observed velocities from SSSE are found to agree well with single-spacecraft (SSEF) analysis techniques applied to the same events. CME propagation directions derived from SSSE and SSEF analysis agree poorly because of known limitations present in the latter. This work was carried out as part of the EU FP7 HELCATS (Heliospheric Cataloguing, Analysis and Techniques Service) project (\url{http://www.helcats-fp7.eu/}).
\end{abstract}
\end{opening}

\section{Introduction}
In addition to the continuous outflow of the solar wind, coronal mass ejections (CMEs; e.g., \citealp{2012Webb}) are a phenomenon by which the Sun releases large quantities of energy in the form of magnetised plasma. They are known to drive magnetic disturbances at Earth, and are in fact the purveyors of the most extreme space weather effects (e.g. \citealp{1991Gosling,2005Huttunen,2012Richardson,2017bKilpua}). CMEs, and their evolution within the solar wind environment, have been the subject of space-based observations since they were first discovered in images taken by the \emph{Orbiting Solar Observatory} 7 (OSO 7, 1971\,--\,74: \citealp{1974Tousey}). Over the following decades, near-continuous coronagraph coverage has been provided by both ground-based instruments, such as the Mauna Loa MK3 Coronameter \citep{1981Fisher}, and their space-based counterparts, most notably the Large Angle and Spectrometric Coronagraph (LASCO, \citet{1995Brueckner}) instruments on board the Solar and Heliospheric Observatory (SOHO). SOHO was launched in 1995 and, despite a brief loss of communication, the LASCO C2 and C3 coronagraphs have operated near-continuously ever since, whilst the inner C1 camera was lost in 1998. The launch of the Solar Mass Ejection Imager (SMEI, 2003\,--\,11: \citealp{2003Eyles}), on the Coriolis spacecraft, extended the coverage of CME observations to far greater solar elongation angles into the heliosphere, by means of wide-angle imaging. Since 2006, the STEREO Heliospheric Imagers (HIs, \citet{2009Eyles}) have continued to provide wide-angle imaging of CMEs. Each STEREO spacecraft possesses two HIs: the HI-1 cameras have an angular range in elongation from 4\,--\,24$^\circ$ and the HI-2 cameras cover 18.7\,--\,88.7$^\circ$, aligned to the ecliptic. Since the launch of STEREO, HI observations, complemented by the STEREO coronagraphs (COR-1 and -2: \citealp{2008Howard}), have provided a considerable amount of information about CME evolution and propagation through the heliosphere (e.g. \citealp{2010Byrne,2010Davis,2010Mostl,2012Savani,2012Harrison,2014Rollett,2014Temmer}). Coronagraphs provide coverage close to the Sun, for example a plane-of-sky (POS) range of 1.1--32R$_\odot$ in the LASCO field of view (FOV), where the POS is the plane that is perpendicular to the Sun--observer line. Conversely, the inner limit of the FOV of the HIs is $4^\circ$ solar elongation (approximately $15$R$_\odot$ in the POS).

Based on observations from LASCO, prior to the loss of C1, and SOHO's Extreme ultraviolet Imaging Telescope (EIT) \citet{2001Zhang,2004Zhang} characterise the acceleration of CMEs into three phases; initiation, impulsive acceleration and propagation. The initiation phase represents the initial acceleration up to approximately $1.3$ to $1.5R_\odot$. Although the  subsequent impulsive acceleration phase of a CME can vary significantly in magnitude and duration, it is typically limited to the inner corona. However, the impulsive acceleration phase can extend throughout the entire LASCO FOV \citep{2000StCyr,2004Zhang}. Typically, beyond a few solar radii, the CME enters its so-called propagation phase. This is characterised by a relatively constant speed, although the very fastest events are seen to exhibit a deceleration, well into the LASCO FOV, and the very slowest an acceleration \citep{2004Yashiro,2009Gopalswamy}. This is evidence of drag forces acting on CMEs and causing their speeds to tend toward the ambient solar wind speed, which is typically 300\,--\,500\,km\,s$^{-1}$. \citet{2017Sachdeva} quantify the contributions from the Lorentz and drag forces for 38 CMEs observed by SOHO and find that the former are most significant between 1.65--2.45\,R$_\odot$, whilst the latter can begin to dominate beyond 4\,R$_\odot$, or up to 50\,R$_\odot$ for slow CMEs.

CMEs can deviate from radial trajectories due to magnetic forces and interaction with background solar wind plasma (e.g. \citealp{2008Kilpua,2010Byrne,2011Lugaz,2015Mostl,2017Manchester}). \citet{2004Cremades} study 276 CMEs observed by LASCO between 1996 and 2002. They find an average latitudinal deflection of $18.6^\circ$ towards the equator during solar minimum (up to 1998) and a poleward deflection of $-7.1^\circ$ during solar maximum (2000), with a period of intermediate behaviour in 1999. \citet{2014Isavnin} studied 14 flux ropes associated with CMEs using multi-viewpoint coronagraph observations combined with MHD modelling to propagate the structures to 1\,AU. Whilst deflection most commonly occurs inside 30\,R$_\odot$, they find that significant deflection, particularly in longitude, can occur out to 1\,AU. Such longitudinal deflections are due to interactions with the background Parker Spiral solar wind structure. \citep{2004Wang} show that this causes faster CMEs to deflect from west to east, whilst slower CMEs deflect from east to west. \citet{2014Isavnin} showed a maximum longitudinal deflection of 29$^\circ$ from the Sun to Earth for an average speed CME, whilst \citet{2014Wang} study an individual interplanetary CME that exhibits a longitudinal deflection of $20^\circ$.

Due to the nature of Thomson scattering of photospheric light by electrons, features observed in coronagraphs are often assumed to be in the POS. This assumption is used to derive POS-projected CME speed and acceleration using observations from a single vantage point because it implies that the position of observed features, along a given LOS, is known. However, due to the wide-angle nature of heliospheric imaging, it is possible to estimate the three-dimensional direction of propagation of a CME using HI data from a single vantage point by assuming that these features are moving at a constant velocity. This is clearly demonstrated in so-called time--elongation maps, or J-maps \citep{1999Sheeley,2008Sheeley}, constructed from HI data, in which this constant linear speed is manifested as an apparent angular acceleration that depends on the propagation direction of the feature with respect to the observing spacecraft. \citet{2009Davies} show that the path of a CME through time--elongation maps may be fitted to retrieve its speed, direction and launch time. Three geometries commonly used to fit CME parameters using a single vantage point are fixed $\phi$ (FP: \citealp{2007Kahler,2008Rouillard,2008Sheeley}), self-similar expansion (SSE: \citealp{2012Davies,2013Mostl}) and harmonic mean (HM: \citealp{2009Lugaz,2010bLugaz}). The geometries model the CME as having a circular cross-sectional front that expands self-similarly with a constant half-width, $\lambda$. The FP and HM models use respective half-widths of $0^\circ$ and $90^\circ$. The SSE geometry is generalised to any half-width and, as such, the FP and HM geometries can be considered as special cases of SSE, where FP is a point source and HM a circle anchored to the Sun. When these geometries are applied to fit CME kinematic properties from time--elongation data, the fitting methods are referred to as FPF, SSEF and HMF. Many studies have shown that CME expansion is indeed close to self similar at interplanetary distances (e.g. \citealp{1997Bothmer,2005Liu,2009Savani}), however, cases of flux ropes that deviate from self-similar expansion are presented by \citet{2012aKilpua,2011Savani,2012Savani}.

In Article 1 \citep{2018Harrison} we present the \textsf{HICAT} (\url{https://www.helcats-fp7.eu/catalogues/wp2_cat.html}) catalogue, which contains a list of all CMEs that were observed using HI (965 by STEREO-A and 936 by STEREO-B) during the science phase of the STEREO mission (April 2007 to September 2014). In Article 2 \citep{2019Barnes}, we present the kinematic properties of these CMEs from the FPF, SSEF and HMF methods, based on single-spacecraft observations from STEREO-A and STEREO-B, which resulted in the generation of the \textsf{HIGeoCAT} (\url{https://www.helcats-fp7.eu/catalogues/wp3_cat.html}) CME catalogue containing 801 and 654 CMEs for STEREO-A and -B, respectively. Here, we take a further subset of these events, which we determine to be CMEs observed in HI images from both spacecraft simultaneously. We apply stereoscopic self-similar expansion (SSSE) geometrical analysis, presented in \citet{2013Davies}, to determine the kinematic properties of these CMEs using observations from both STEREO spacecraft. The SSSE method is based on the SSE geometry. SSSE with $\lambda=0^\circ$ (i.e. a point source) corresponds to the so-called geometric triangulation (GT) technique, first performed by \citet{2010Liu}. SSSE with $\lambda=90^\circ$ (i.e. a circle anchored to the Sun) corresponds to the tangent to a sphere (TAS) technique, introduced by \citet{2010aLugaz}. The extra information afforded by using a second vantage point allows one to drop the assumption that a CME is travelling in a constant direction and at a constant speed. However, the CME is still assumed to be self-similarly expanding at a constant half-width. These sterescopic methods may only be applied to features that propagate in the plane containing both observing spacecraft and the Sun, which, in the case of STEREO, is the ecliptic.

\citet{2010bLugaz} analysed 12 CMEs that occurred between 2008 and 2009 and were seen in HI on both STEREO spacecraft using single-spacecraft and stereoscopic methods based on the FP ($\lambda=0^\circ$) and HM ($\lambda=90^\circ$) geometries. For each CME, they found poor correlation between the propagation direction derived from observations using STEREO-A and STEREO-B, when each spacecraft was used individually to track the CMEs. This discrepancy was, however, smaller when the larger half-width was used. They identified three main sources of error: the assumption of constant propagation direction when using observations from a single-spacecraft, the assumption of negligible width when using $\lambda=0^\circ$ and the assumption of constant CME velocity. The authors show that the first and third of these may be addressed by using stereoscopic observations, whilst the second may by addressed by modelling the CMEs with non-zero half-width. The single-spacecraft methods, and their ability to predict arrival times, are assessed by \citet{2011Mostl}, who find arrival times within $\pm$5 hours can be achieved if CMEs are tracked to at least 30$^\circ$ elongation. \citet{2017Mostl} study the \textsf{HiGeoCAT} CMEs, whose kinematic properties are derived by the SSEF method (using $\lambda=30^\circ$), and their ability to predict arrival times at different spacecraft. They find a range of 23-35\% of predicted arrivals are actually observed in situ at the various spacecraft. The predicted arrival times are early by an average of 2.6 $\pm$ 16.6 hours, excluding predicted and in-situ signatures that lie outside a 1\,day time window. More sophisticated geometries for modelling CME morphology include the Graduated Cylindrical Shell (GCS) model \citep{2006Thernisien}, ElEvoHI \citep{2016Rollett}, which employs an elliptical CME front, and includes the effects of drag, and 3DCORE \citep{2018Mostl}, which is used to measure CME rotation and deflection and also to give information about magnetic field orientation based on data from the Solar Dynamics Observatory. Both \citet{2015Volpes} and \citet{2019Palmerio} apply SSSE analysis to HI data, where the half-width of the CME is first determined using coronagraph observations. This is rather labour intensive and so is more challenging to apply to large statistical studies, such as that which we present here.

These single-spacecraft (SSEF, including FPF and HMF) and stereoscopic (SSSE, including GT and TAS) analysis techniques are based on assumptions that often fail to include the more complex physical processes that occur during CME propagation, such as rotations (e.g. \citealp{2008Mostl,2011Vourlidas,2018Mostl}), deformations (\citealp{2010Savani}) and, for single-spacecraft techniques in particular, deflections (e.g. \citealp{2010Byrne,2014Wang}). These effects result from CMEs interacting with other structure in the heliosphere including high speed streams and  other CMEs \citep{2009Lugaz,2012Lugaz,2014Liu,2017Lugaz}. Whilst CME\,--CME interactions may be quite rare during solar minimum, a CME rate of 5\,--10\,day$^{-1}$ is not unusual at solar maximum (e.g. \citet{2004Yashiro,2009Robbrecht,2010Gopalswamy,2017Vourlidas,2018Harrison}). Indeed, \cite{2007Zhang} show that of the 88 major geomagnetic storms (defined by minimum disturbance storm time index, $Dst \leq$\,100nT) that occurred during solar cycle 23, 60\% were associated with individual CMEs, whilst 27\% were the result of CME interactions with background structure or other CMEs.

Whilst the STEREO mission comprises two spacecraft, contact with STEREO-B was lost in October 2014. The recently launched Parker Solar Probe and Solar Orbiter missions both possess wide-angle imagers (Wide-Field Imager for Parker Solar Probe, or WISPR: \citealp{2016Vourlidas} and SoloHI: \citealp{2013Howard}, respectively), as will the upcoming PUNCH mission in low Earth orbit and the potential future mission to the Sun-Earth L5 point \citep{2017Kraft}. These newly-launched spacecraft are already returning CME images (\citealp{2020Hess}, using WISPR) and it is therefore important to realise the benefits and the limitations of the methods that we use to analyse the data, particularly the assumptions involved when observing from just a single vantage point.

Section \ref{sec:method} includes a description of the SSSE method and an explanation of how we apply it to time--elongation profiles from HI. Section \ref{sec:results} presents the results of the statistical analysis of the SSSE fitting results, including CME acceleration. Finally, we present a comparison between the kinematic properties derived using stereoscopic analysis methods and those that were determined using observations from just a single spacecraft. For a thorough description the compilation of \textsf{HICAT} the reader is urged to refer to Article 1, whilst the compilation of \textsf{HiGeoCAT} is the subject of Article 2.

\section{Method}
\label{sec:method}

\begin{figure}
  \centering
  \includegraphics[width=0.49\textwidth]{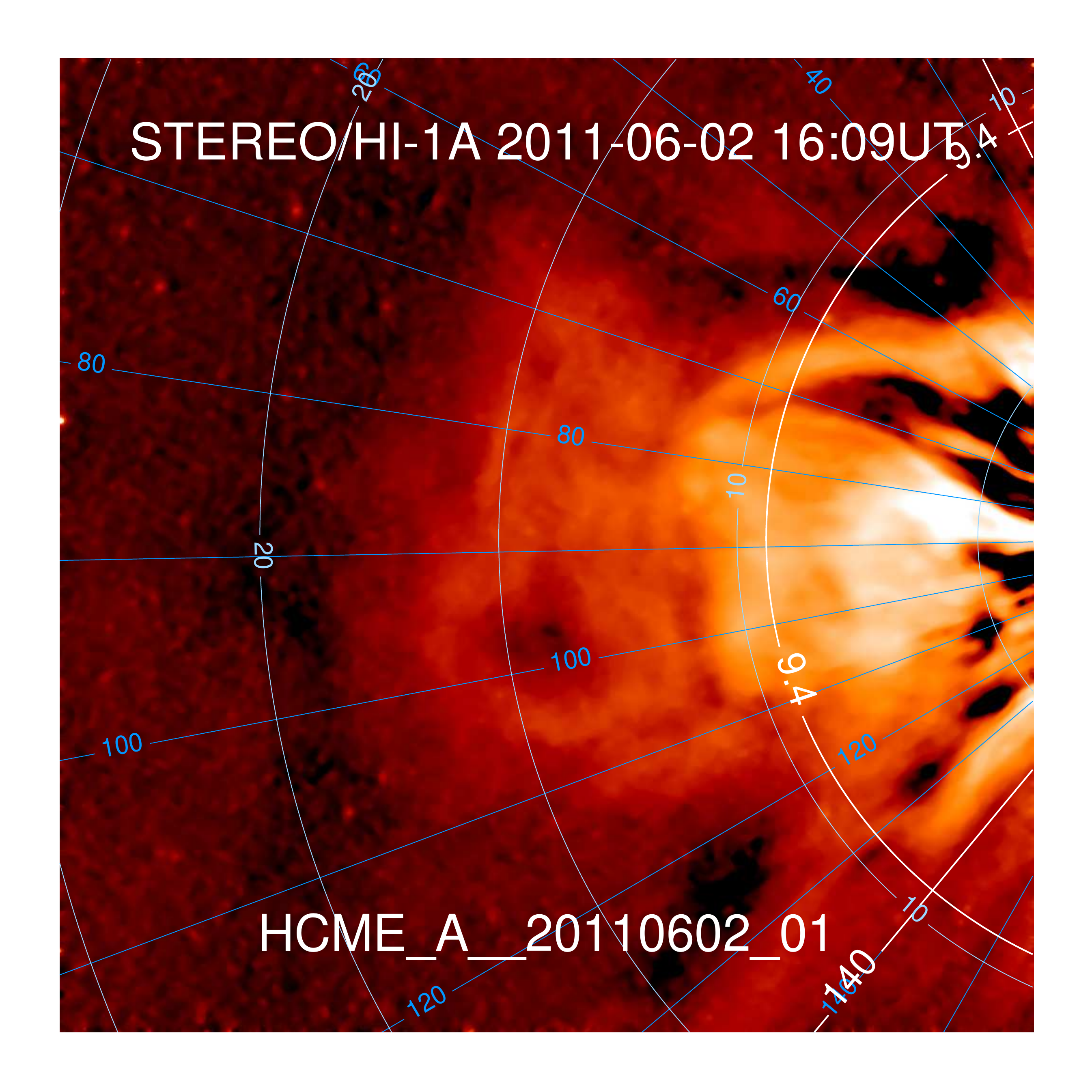}
  \includegraphics[width=0.49\textwidth]{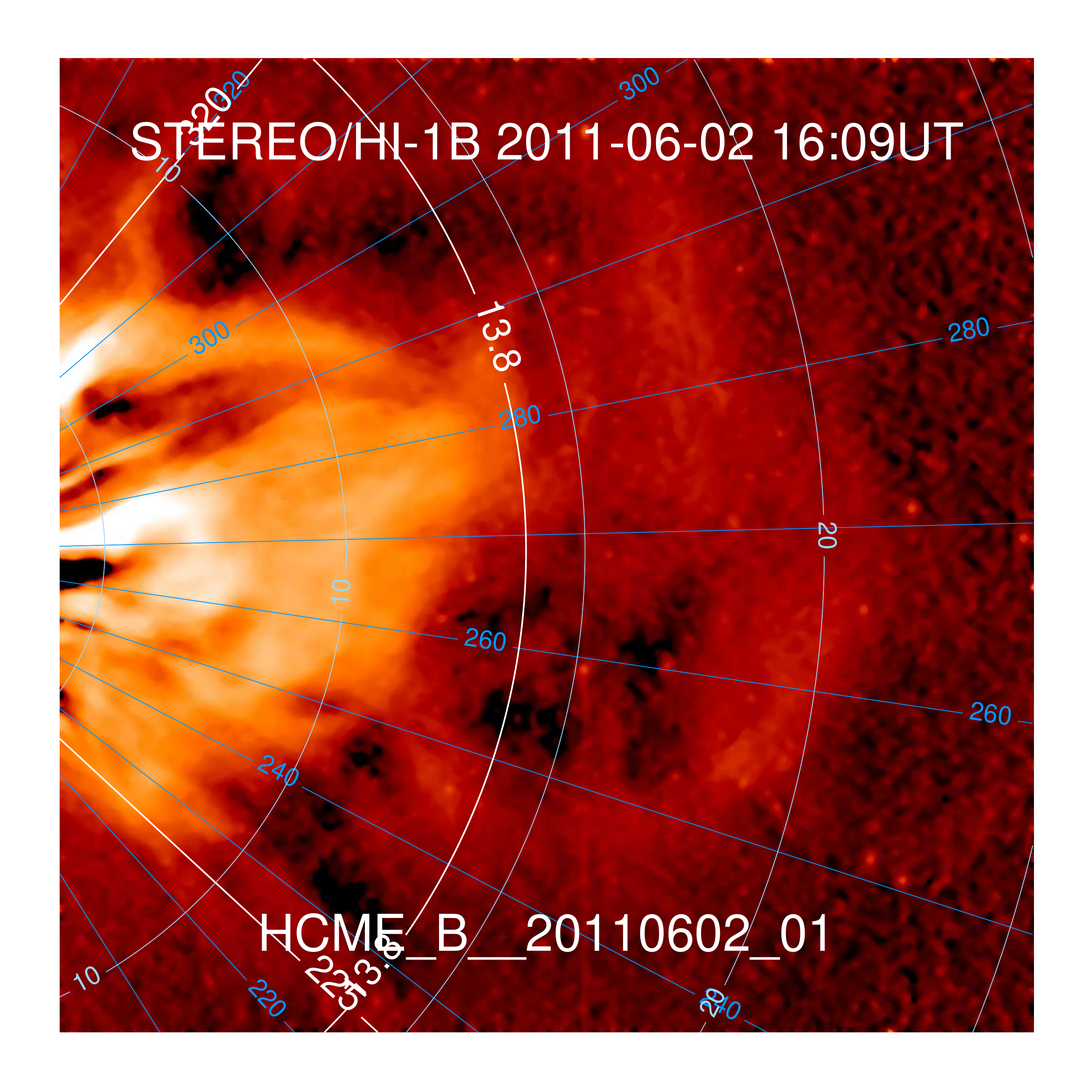}
  \includegraphics[width=0.49\textwidth]{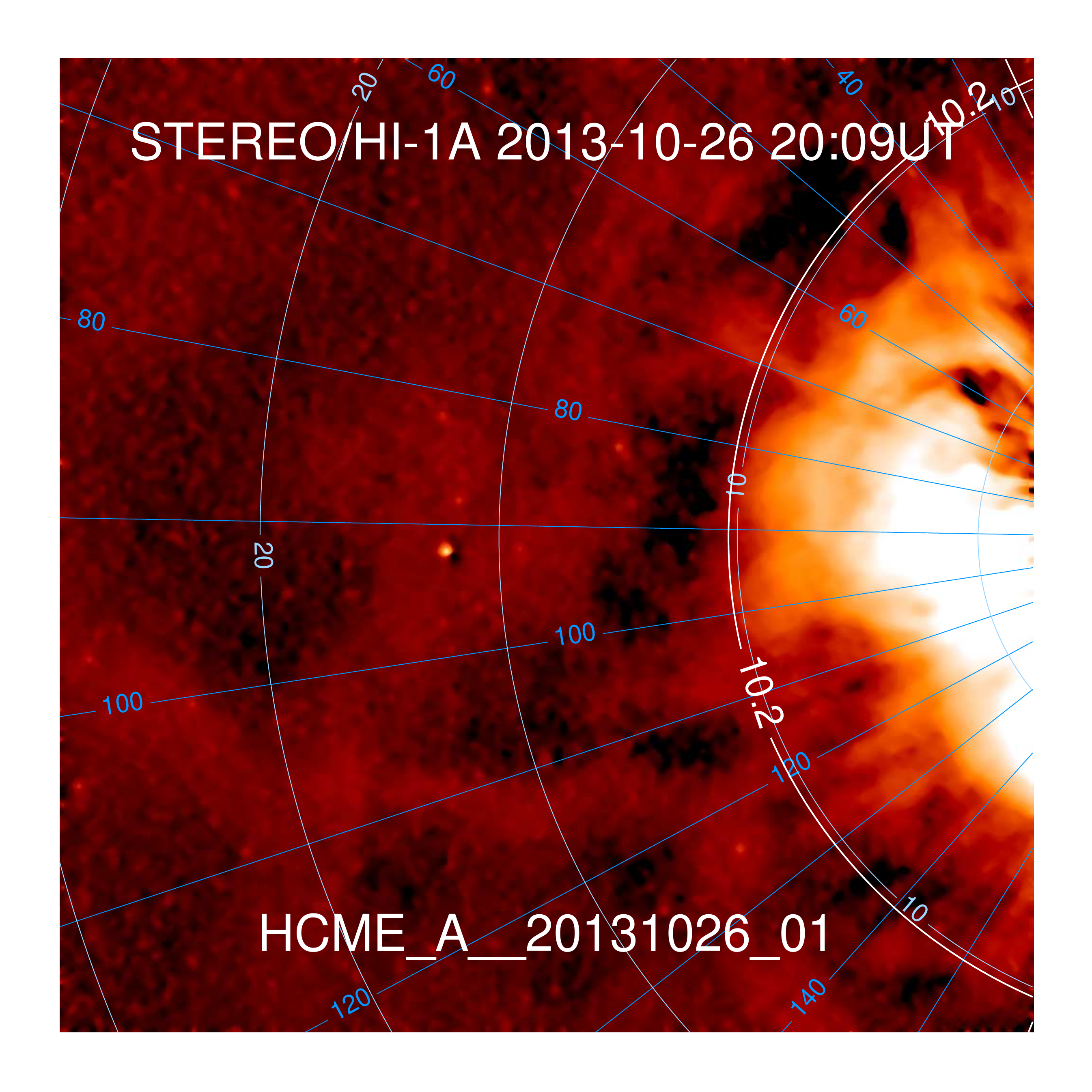}
  \includegraphics[width=0.49\textwidth]{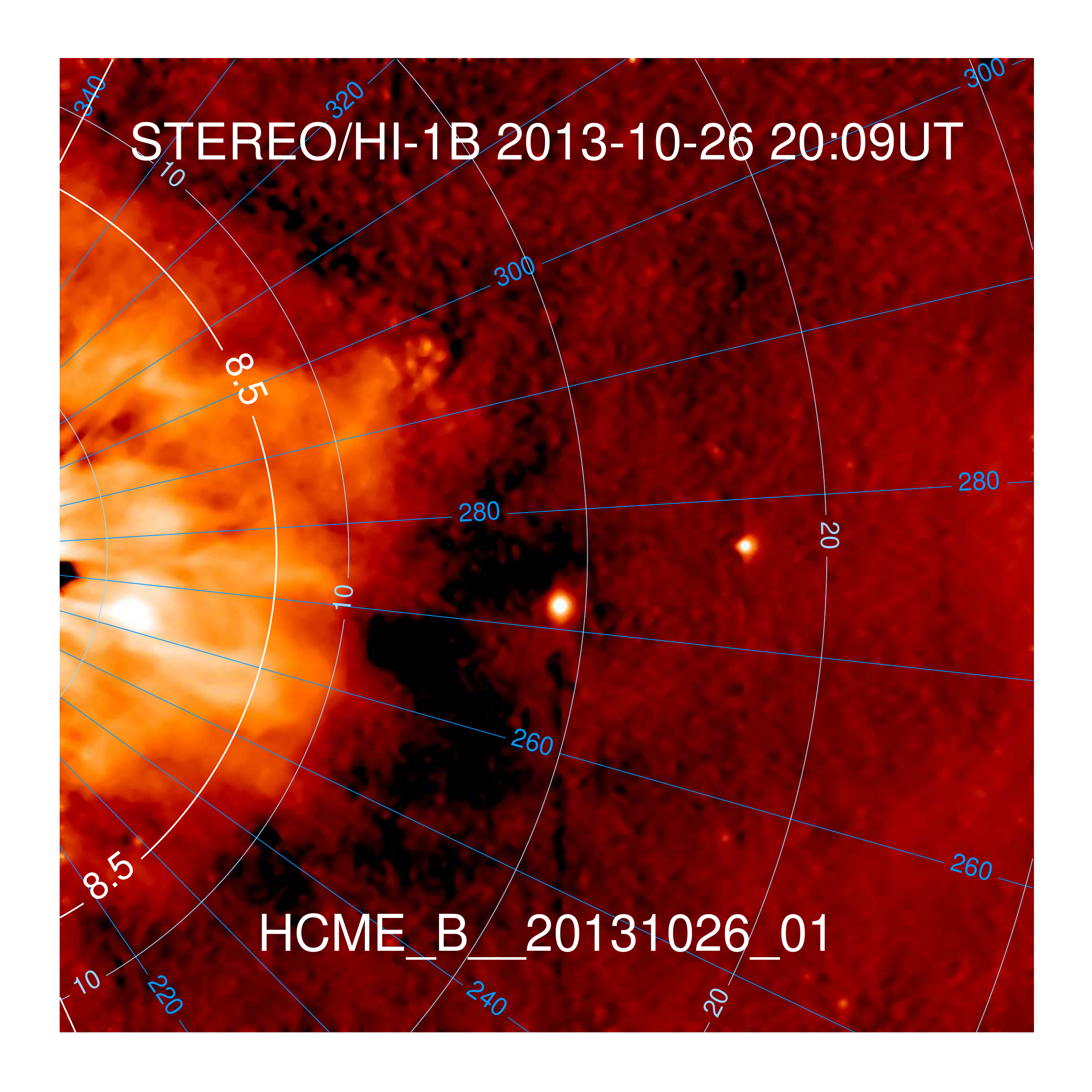}
  \caption{Background subtracted images of two example \textsf{HIJoinCAT} CMEs, which were observed by HI-1 on both STEREO-A and -B. Panels \textbf{a} and \textbf{b} show a CME that occurred in June 2011 (\textsf{HCME\_A\_\_20110602\_01} or \textsf{HCME\_B\_\_20110602\_01}) when the STEREO spacecraft were separated by approximately 192$^\circ$. Panels \textbf{c} and \textbf{d} show a CME that occurred in October 2013 (\textsf{HCME\_A\_\_20131026\_01} or \textsf{HCME\_B\_\_20131026\_01}) when the spacecraft were on the far-side of the Sun, each close to 145$^\circ$ from Earth. 5$^\circ$ contours of elongation and position angle are over-plotted in helioprojective coordinates. In each case the upper and lower PA extent of the CME is also over-plotted, as is the approximate elongation of the leading edge.}
\label{fig:img}
\end{figure}

\begin{figure}
\centering
\includegraphics[width=\textwidth]{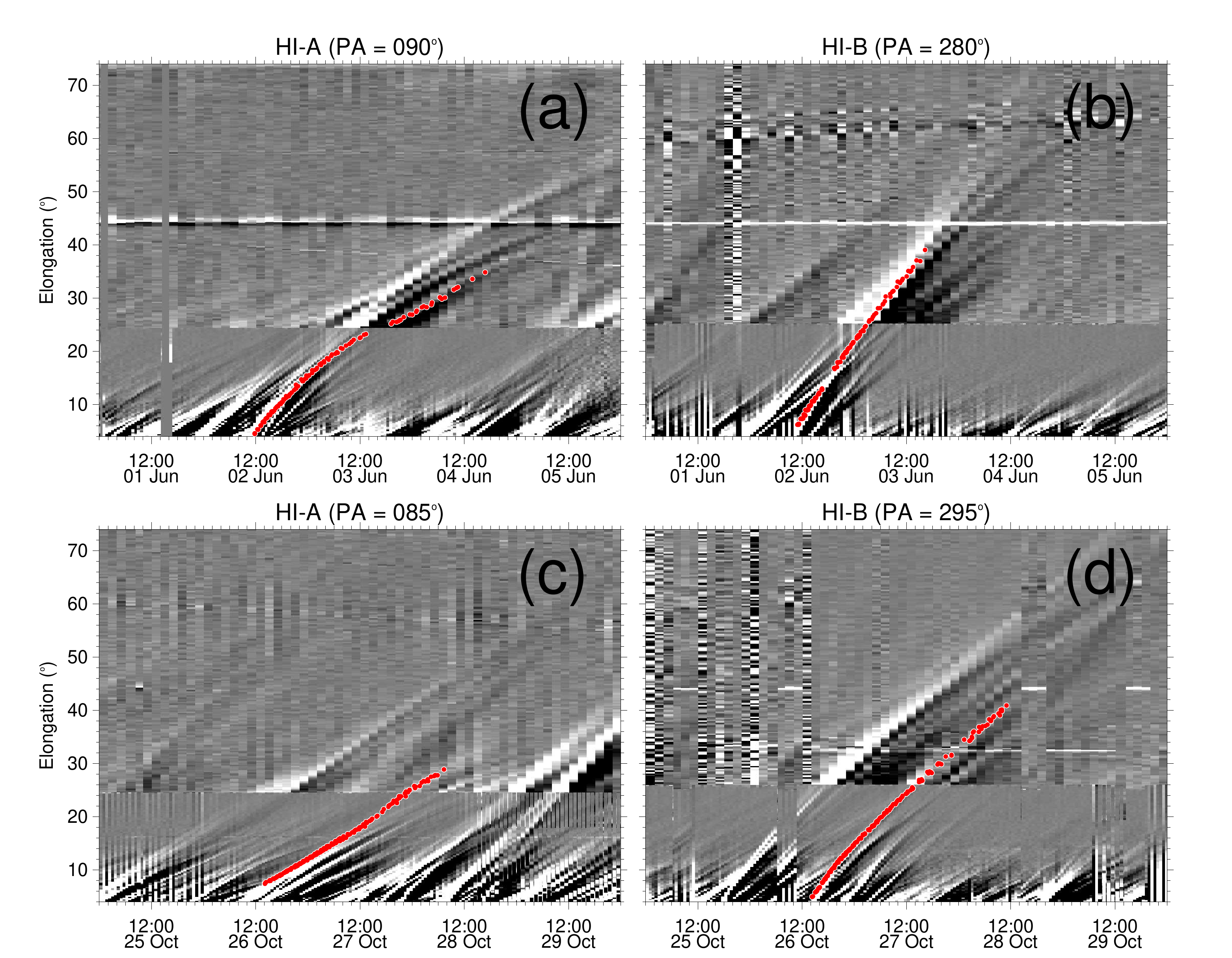}
\caption{Time--elongation maps created by stacking slices of HI-1 and HI-2 difference images, taken from a fixed position-angle, in time. The red data over-plotted in each panel correspond to the observed leading edge of the CME. Panels \textbf{a} and \textbf{b} show \textsf{HCME\_A\_\_20110602\_01} and Panels \textbf{c} and \textbf{d} show \textsf{HCME\_A\_\_20131026\_01}, which correspond to the CMEs shown in figure \ref{fig:img}.}
\label{fig:jmap_figure}
\end{figure}

Due to the overlap of the FOVs of the HI cameras on the two STEREO spacecraft, a number of CMEs occur that are imaged from both STEREO-A and -B vantage points. The amount by which the FOVs of each spacecraft overlap increased from the start of the mission until late 2010, when the spacecraft separation was close to 180$^\circ$, after which time it progressively decreased. This means that it is typically easier to identify CMEs that are observed by HI on both spacecraft near the end of 2010. Additionally, the CME rate was very low at the start of the mission because it coincided with solar minimum (Article 1 and 2). As a result of these two factors, the first event that we identify to be observed in HI on both spacecraft is in August 2008, nearly two years after the launch of STEREO.

We identify joint events by manual inspection of those events contained in \textsf{HICAT}, based upon their time of entry into the respective HI-1 FOVs. If a \textsf{HICAT} CME enters the HI-1A FOV, we identify if any CMEs have entered the HI-1B FOV within a $\pm$2\,day window. This time window is chosen to be large enough that no CME that is observed by both spacecraft is likely to lie outside it. We then determine if the two events are the same CME by examination of sequences of simultaneous images from HI-1A and HI-1B, which have a nominal cadence of 40\,minutes. This is achieved by identifying similarities in CME size, morphology and internal structure. The events that satisfy these conditions are listed in a new, separate catalogue that is available on the HELCATS website (\textsf{HIJoinCAT}: \url{https://www.helcats-fp7.eu/catalogues/wp2_joincat.html}). \textsf{HIJoinCAT} contains just two columns, which list the unique identifier of the CME observed in HI-1 data from STEREO-A and STEREO-B, respectively, as they appear in the single-spacecraft \textsf{HICAT} list (and therefore \textsf{HIGeoCAT}, of which the former is a super-set). \textsf{HiJoinCAT} contains a total of 273 CMEs, observed between 31 August 2008 and 02 April 2014. Unlike both \textsf{HICAT} and \textsf{HIGeoCAT}, the \textsf{HIJoinCAT} list is no longer being updated due to the fact that it requires data from both spacecraft, a condition that is no longer satisfied since the loss of STEREO-B in 2014.

Two \textsf{HIJoinCAT} CMEs are shown in Figure \ref{fig:img}; the \textsf{HICAT} unique identifier for each event is printed at the bottom of each page. The top two panels show simultaneous images from HI-1A (left) and HI-1B (right) at 16:09UT on 02 June 2011 when STEREO-A and STEREO-B were 192$^\circ$ apart. We are able to determine quite easily that these images are of the same CME because they exhibit very similar structure. The bottom two panels in Figure \ref{fig:img} show simultaneous images from HI-1A (left) and HI-1B (right), at 20:09UT on 26 October 2013, when the STEREO spacecraft were 290$^\circ$ apart. Although it is less obvious than in the previous example, similar structures can still be identified in each image, which leads us to conclude that this is indeed the same CME observed by both spacecraft.

\begin{figure}
\centering
\includegraphics[width=\textwidth]{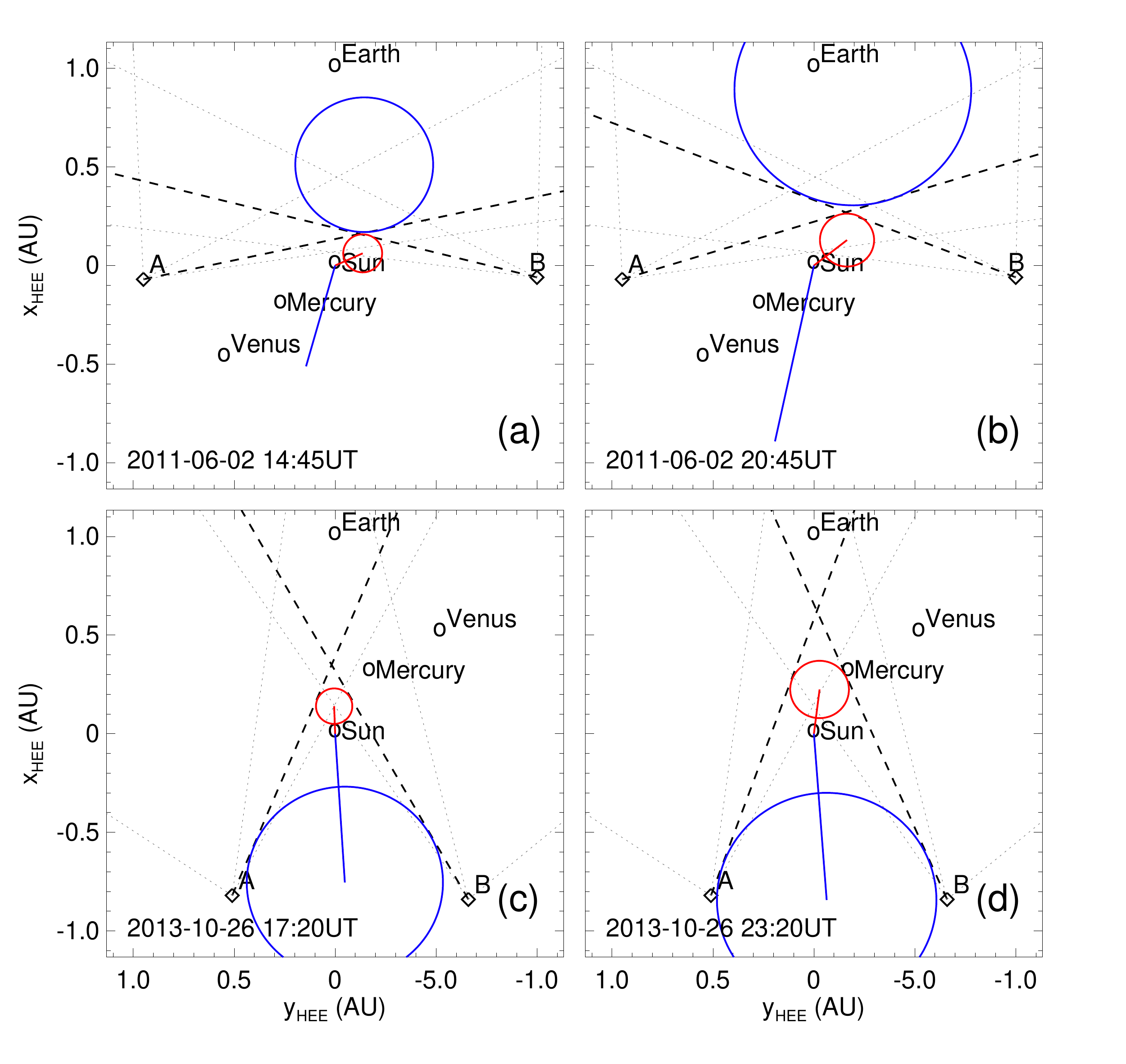}
\caption{Schematic showing example solutions to Equation \ref{eq:ssse} for two $\lambda=40^\circ$ CMEs in the ecliptic plane using the Heliocentric Earth Ecliptic (HEE) coordinate system. Panels \textbf{a} and \textbf{b} show \textsf{HCME\_A\_\_20110602\_01}/\textsf{HCME\_B\_\_20110602\_01} and Panels \textbf{c} and \textbf{d} show \textsf{HCME\_A\_\_20131026\_01}/\textsf{HCME\_B\_\_20131026\_01}, the same CMEs shown in Figures \ref{fig:img} and \ref{fig:jmap_figure}. STEREO-A is labelled $A$ and STEREO-B is labelled $B$. The dashed black lines radiating from each spacecraft are the measured elongation of the CME front and the dotted grey lines delimit the HI-1 and HI-2 FOVs. Panels \textbf{a} and \textbf{b} show solutions separated by six hours, as do Panels \textbf{c} and \textbf{d}. In all panels the red circle represents the 'correct' solution, whilst the blue circle is the solution that we discard.}
\label{fig:ssse_cartoon}
\end{figure}

To the CMEs that are contained in the new catalogue, we apply the SSSE analysis method of \citet{2013Davies} to the time--elongation data that were already determined for each CME front for \textsf{HIGeoCAT} as described in Article 2. It should be noted that the time--elongation data of each CME front were determined at a position angle (PA) close to the apex, which is occasionally far from the ecliptic. The SSSE fitting method must be applied to a CME front in the plane that contains both spacecraft and the Sun, in the case of STEREO this is the ecliptic plane. As such, we must assume that those time--elongation profiles recorded away from the ecliptic are a reasonable approximation to the time--elongation profile of the CME front in the ecliptic plane. Examples of CME time--elongation maps are shown in Figure \ref{fig:jmap_figure}, where panels a and b correspond to the CME in the top panels of Figure \ref{fig:img}. Likewise, panels c and d of Figure \ref{fig:jmap_figure} correspond to the CME shown in the bottom of Figure \ref{fig:img}. The SSSE method makes the same assumptions about CME morphology as does the SSEF method \citep{2012Davies}: that the CME is assumed to be a self-similarly expanding structure, with a specified half width and a circular cross-sectional front. However, unlike with the single-spacecraft methods, the extra information afforded by using two spacecraft to track the CME means that we no longer require the assumptions of constant speed and constant propagation direction that were necessary with the single-spacecraft fitting technique. By applying the SSSE method to a CME, using an angular half-width, $\lambda$, its position is defined by the points where the line of sight from each spacecraft intercepts its leading edge. Assuming a circle of fixed half-width, we determine the heliocentric distance of the CME apex, $R$, using the following equation from \citet{2013Davies}

\begin{equation}
R=\frac{d_A\sin(\epsilon_A(t))(1+\sin(\lambda))}{\sin(\epsilon_A(t)+\phi_A)+\sin(\lambda)}
=\frac{d_B\sin(\epsilon_B(t))(1+\sin(\lambda))}{\sin(\epsilon_B(t)+\phi_B)+\sin(\lambda)}
\label{eq:ssse}
\end{equation}

\noindent
where $d$ is the heliocentric distance of the spacecraft, $\epsilon$ is the solar elongation angle of the CME leading edge and $\phi$ is the spacecraft-Sun-CME apex angle. The subscripts A and B refer to the observing spacecraft. From Equation \ref{eq:ssse}, the time--elongation profiles that were used to compile \textsf{HIGeoCAT} are used to solve for $R$ and $\phi$, as a function of time. For a full mathematical derivation, the reader is referred to \citet{2013Davies}, however Figure \ref{fig:ssse_cartoon} illustrates the concept for our two example CMEs.

Figure \ref{fig:ssse_cartoon} shows a schematic representation of the solutions for two CMEs with $\lambda=40^\circ$ that were derived for two time-steps in the time--elongation profiles for the CMEs shown in Figures \ref{fig:img} and \ref{fig:jmap_figure}. The method for solving Equation \ref{eq:ssse} for $R$ and $\phi_A$ or $\phi_B$ \citet{2013Davies} depends on a square root and therefore has two solutions. Typically, however, one of these solutions is unphysical and may be easily discarded. Mathematically, the blue circle in panels a and b of Figure \ref{fig:ssse_cartoon} describes a CME propagating away from the Sun with a negative $R$, of which the trailing edge corresponds to the observed elongation. This is why the blue line does not connect from the Sun to the CME centre. In these such cases, we can easily discard this solution as incorrect. Conversely, the red circle represents the leading edge of a CME travelling approximately between the two spacecraft, which is consistent with the observations in Figure \ref{fig:img}. In some cases there exist two ambiguous realistic solutions and the appropriate result must be selected manually, for example the second row in figure \ref{fig:ssse_cartoon}. In these cases, we assume the CME to be directed approximately towards the Earth because this is the region where the HI-1 FOVs overlap. There exists a further limitation of stereoscopic methods when the observed lines of sight are close to parallel, that is, when the CME leading edge passes directly between the two spacecraft. When this happens, $R$ and $\phi$ become strongly influenced by small errors in $\epsilon$ and the resulting solutions give CMEs that vary significantly in apex position between successive observations. For this reason we discard these solutions from the analysis presented in this article, however these CMEs are still included in \textsf{HIJoinCAT}, which does not contain CME kinematics. This configuration is most common when the spacecraft are separated by approximately 180$^\circ$, which is close to solar maximum and when the overlap between the HI FOVs is also maximised, and therefore the time at which the majority of the joint CMEs are detected.

\begin{figure}
\centering
\includegraphics[width=\textwidth]{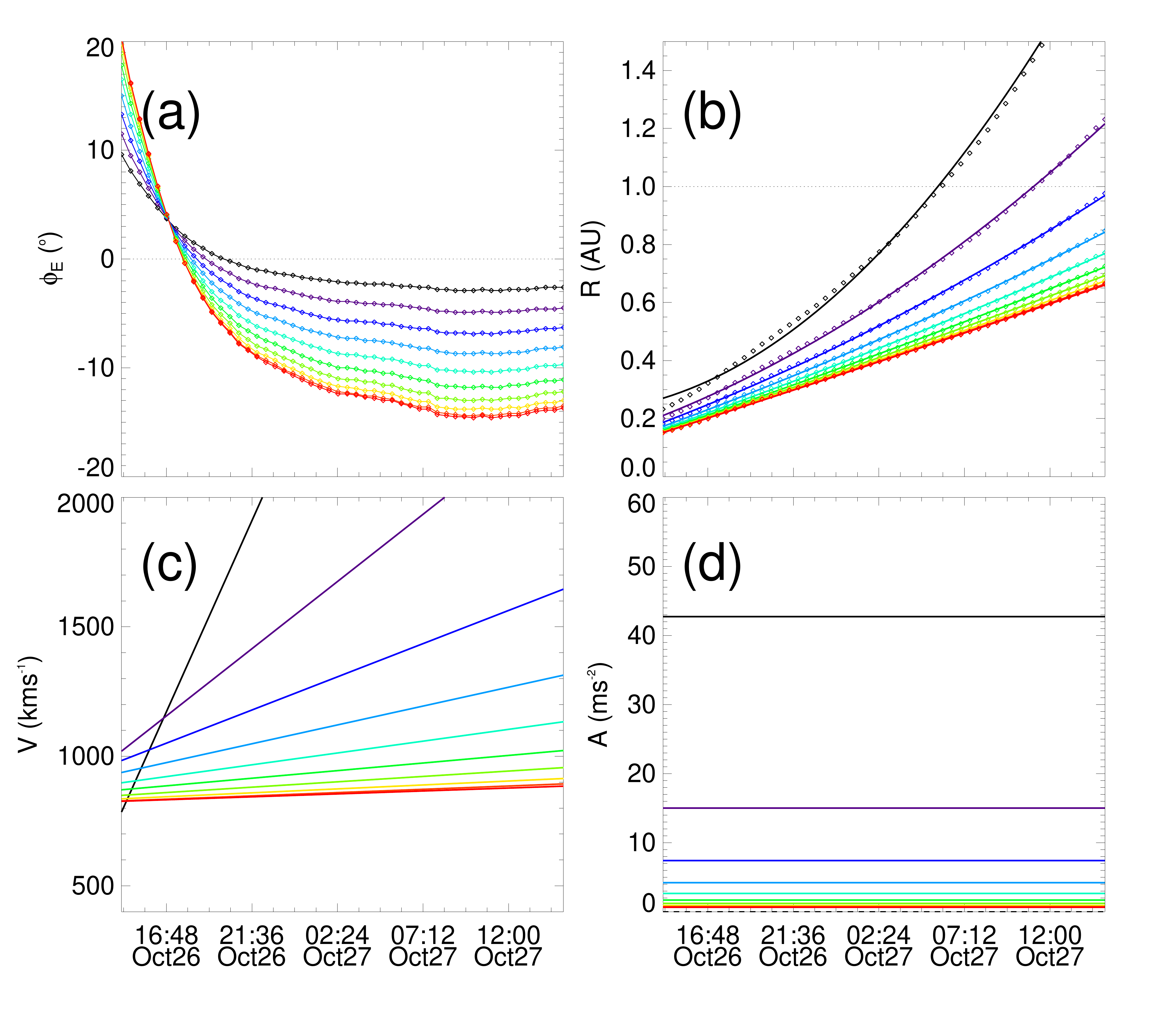}
\caption{SSSE-derived kinematic properties for the CME observed on 26 October 2013. The procedure is applied using ten different half-widths in Equation \ref{eq:ssse}, increasing from $0^\circ$ (black profiles) to 90$^\circ$ (red profiles) in steps of 10$^\circ$. \textbf{a} shows the CME apex longitude in the ecliptic plane relative to the Earth. \textbf{b} shows the CME apex distance from the Sun as a function of time. \textbf{c} and \textbf{d} show CME velocity and acceleration, respectively. The latter are calculated by taking the first and second derivatives of a second order polynomial fit to $R$, with respect to time. These fits are over-plotted in \textbf{b} for each $10^\circ$ value of $\lambda$, from $\lambda=0^\circ$ (black) to $\lambda=90^\circ$ (red).}
\label{fig:ssse_kinematics}
\end{figure}

The image cadences of the HI-1 and HI-2 cameras are 40 minutes and 2 hours, respectively, and so we may use successive sets of observations to track the time--elongation profile of the CME's leading edge, using J-maps, as it propagates through the heliosphere, as shown in Figure \ref{fig:jmap_figure}. The SSSE technique requires that the elongation of the CME front as observed from STEREO-A and -B be simultaneous, so the time--elongation profile from each spacecraft is linearly interpolated onto a set of common times, separated by 30 minutes, limited by the time interval for which data from both vantage points are available. For a given time-step,  a value of $R$ and $\phi$ is calculated using equation \ref{eq:ssse} for ten different half-widths increasing from 0$^\circ$ to 90$^\circ$ in increments of 10$^\circ$. Such analysis is performed for each time-step to produce time profiles of $R$ and $\phi$ for the CME. To derive the profiles of the CME velocity, $V(t)$, and acceleration, $A$, we fit a function to the CME apex radial distance ($R$) profile. As is the case with many existing CME catalogues (e.g. \citet{2004Yashiro,2017Vourlidas}), we choose to fit a second order polynomial.

An example of the analysis of $HCME\_A\_\_20131026\_01$ and $HCME\_B\_\_20131026\_01$ is shown in Figure \ref{fig:ssse_kinematics}, where the CME is tracked for just over 24 hours (50 half-hour time-steps). The CME apex longitude, in Heliocentric Earth Ecliptic (HEE) coordinates, as a function of time (panel a) can be seen to shift from approximately +20$^\circ$ to -15$^\circ$, in the most extreme case of $\lambda=90^\circ$, over this period, where positive is westward. A deflection of this magnitude is feasible \citep{2014Wang,2014Isavnin}, and the west-to-east direction is consistent with the findings of \citet{2004Wang} for fast CMEs. However, we expect that some contribution is likely to result from errors in the fitting method. \citet{2013Liu} find the HM geometry to be an inaccurate approximation for CMEs near the Sun due to the fact that CMEs expand at a rate greater than self-similarity in their early propagation phase. Indeed, the deflection shown in Figure \ref{fig:ssse_kinematics} is most pronounced for $\lambda=90^\circ$ and least so for $\lambda=0^\circ$. Panel b of Figure \ref{fig:ssse_kinematics} shows the CME apex heliocentric distance, in AU, as a function of time. The first time-step corresponds to a CME apex distance close to 0.2\,AU, relatively independent of $\lambda$, and, depending on the chosen half-width, the CME is tracked to just beyond 0.6\,AU (red line; $\lambda=90^\circ$) or well beyond 1.5\,AU (black line; $\lambda=0^\circ$). For each half-width, the second-order polynomial fitted to the $R$ profile is over-plotted as a solid line. The velocity profile, in km\,s$^{-1}$, derived from this second-order fit is shown in panel c, as is the acceleration, in panel d. The fit to the 0$^\circ$ half-width CME suggests an acceleration rate of over 40\,m\,s$^{-2}$, resulting in a speed that increases from 800\,km\,s$^{-1}$ to well over 2000\,km\,s$^{-1}$ in less than eight hours. Such a speed increase is inconsistent with typical CME behaviour, particularly at these radial distances, which suggests that using this half-width to model the CME is a poor approximation. Indeed, for the 90$^\circ$ half-width model we find a CME accelerating at approximately 1\,m\,s$^{-2}$, maintaining a speed between 800\,--\,900\,km\,s$^{-1}$, which, although fast, is certainly more realistic behaviour. This example illustrates an important result that is common to many of the CMEs analysed in this article: using a small half-width often results in unphysical CME acceleration to very high velocities. This is unrealistic given drag from the background solar wind is the mechanism by which CMEs are expected to change speed this far from the Sun.

\section{CME Statistical Properties}
\label{sec:results}

\subsection{CME Frequency}

\begin{figure}
\centering
\includegraphics[width=\textwidth]{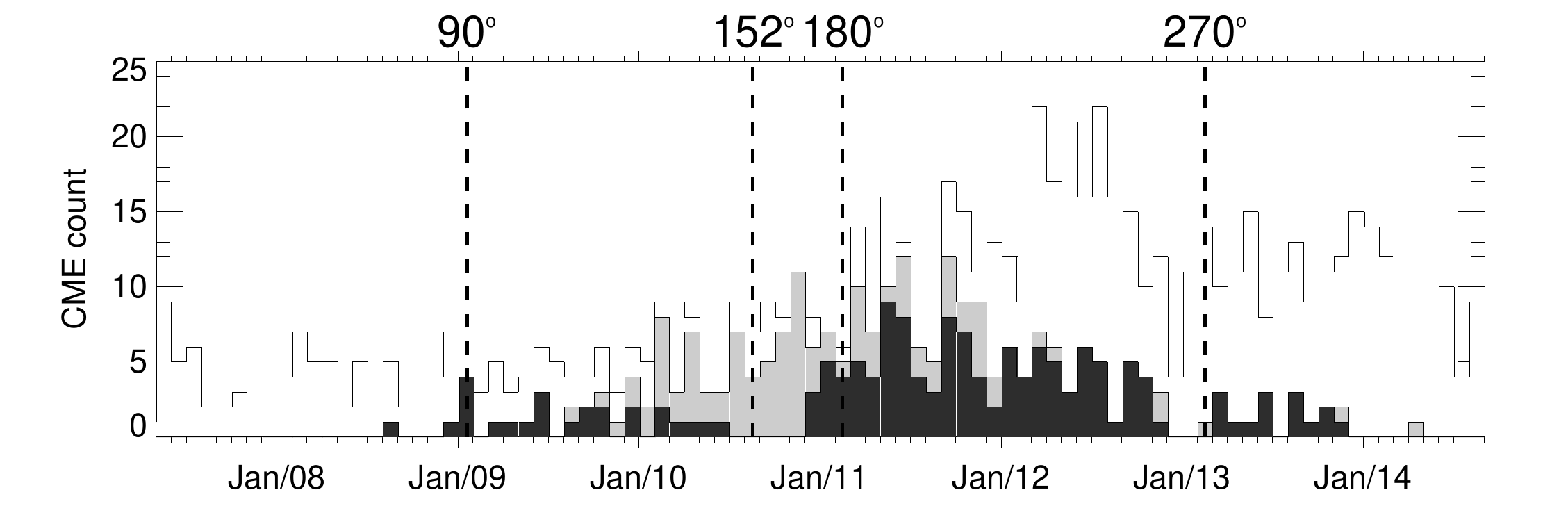}
\caption{Histogram showing monthly CME count during the period April 2007 to September 2014. The total bin height (white) shows the highest number of CMEs detected by either STEREO-A or -B for a given month from \textsf{HICAT}. The total height of all shaded regions represents CMEs that were identified to be present in both HI-1A and HI-1B images (i.e. the CMEs in \textsf{HIJoinCAT}), whilst the darkest region is the number of CMEs to which the stereoscopic fitting has been applied. The light grey region shows CMEs that were omitted from the stereoscopic fitting. The vertical dashed lines represent the evolution of the spacecraft separation angle during this interval, where 152$^\circ$ is when the boresights of the two HI-1 cameras are directly aligned.}
\label{fig:stereo_hist}
\end{figure}

Initially we identify CMEs in HICAT that are observed using both STEREO-A and -B. For each CME observed in HI-1A images we identify any CMEs that enter the HI-1B FOV within $\pm$2\,days of the time that the CME enters the HI-1A FOV. For all \textsf{HICAT} CMEs (965 from STEREO-A and 936 from -B) observed prior to the loss of communication with STEREO-B, we produced a preliminary list of 475 potentially common CMEs using this method. We refine this list through examination of HI-1 images. This results in a subset of 273 CMEs imaged by both HI-1A and HI-1B for inclusion in our so-called \textsf{HIJoinCAT} catalogue. It is likely that we erroneously exclude some events that were actually observed by both spacecraft due to non-optimal viewing geometry. To these 273 CMEs, we apply the aforementioned stereoscopic fitting analysis to the STEREO-A and STEREO-B time--elongation profiles from the \textsf{HIGeoCAT} catalogue. Figure \ref{fig:stereo_hist} shows the temporal distribution of the CME count with a bin size of one month. The white bins show the greatest number of CMEs observed by in HI on either STEREO-A, or -B, from \textsf{HICAT}. The shaded regions show the total number of \textsf{HIJoinCAT} CMEs, which is greatest during 2010 and 2011, corresponding to the time when the spacecraft were close to $180^\circ$ separation; solar cycle 24 peaked soon after, in 2012. The lighter grey region of the histogram, shows the number of CMEs that were confirmed to be imaged by both HI-1A and HI-1B but which were excluded from the final analysis for one of two reasons. Firstly, we exclude some CMEs for which time--elongation profiles from both STEREO-A and STEREO-B view points are not available in the \textsf{HIGeoCAT} catalogue. This is due to data gaps or CMEs that were too difficult to track. Secondly, and more significantly, the SSSE method breaks down when the LOSs of the observed leading-edge of the CME from both spacecraft are approximately parallel. This occurs commonly when the boresights of the HI-1 cameras are directly opposite, because the majority of CMEs were found not to be tracked far into the HI-2 FOV in Article 2. As the HI-1 FOVs are centred at $14^\circ$ elongation in the ecliptic, this alignment occurs around August 2010, when the spacecraft are separated by $152^\circ$. Figure \ref{fig:stereo_hist} shows the result of this problem, where all 35 dual-spacecraft CMEs observed in the months July-November 2010 are excluded from the final analysis. The majority of CMEs that were analysed using SSSE is greatest during 2011 and 2012, which is when the spacecraft separation approaches $270^\circ$ and coincides with solar maximum. In total, the stereoscopic analysis was successfully applied to 151 CMEs.

\subsection{CME Acceleration and Deflection}

\begin{figure}
\centering
\includegraphics[width=\textwidth]{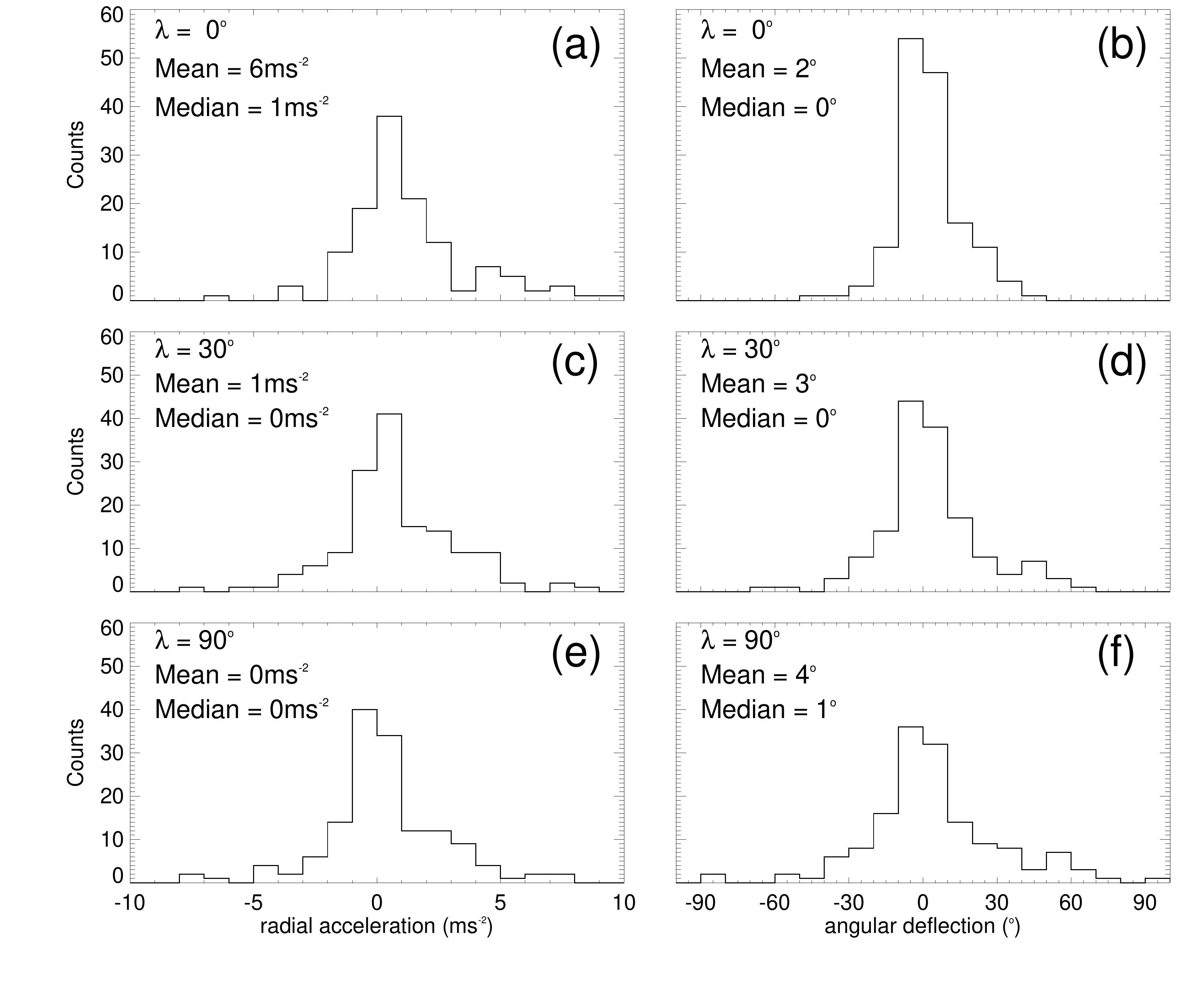}
\caption{Histograms showing the distributions of CME accelerations (panels \textbf{a}, \textbf{c} and \textbf{e}) calculated with a bin-width of 1\,m\,s$^{-2}$ and angular deflections (panels b, d and f) calculated with a bin-width of $10^\circ$. Panels \textbf{a} and \textbf{b} use $\lambda=0^\circ$, panels \textbf{c} and \textbf{d} use $\lambda=30^\circ$ and panels \textbf{e} and \textbf{f} use $\lambda=90^\circ$. The acceleration values are determined by fitting a second-order polynomial to the CME apex distance profile. CME deflection represents the total change in angular position of the CME apex, between the first and last interpolated time-step. In panels \textbf{a}, \textbf{c} and \textbf{e} 23, 5 and 3 CMEs, respectively, exceed the upper 10\,m\,s$^{-2}$ limit of the plot.}
\label{fig:deflection}
\end{figure}

Figure \ref{fig:deflection} shows the distributions of both CME acceleration (panels a, c and e), which is assumed to be constant, and CME longitudinal deflection (panels b, d and f), determined using SSSE with half-widths of 0$^\circ$ (top row), 30$^\circ$ (middle row) and 90$^\circ$ (bottom row). The acceleration distributions are peaked near zero, showing that, typically, CMEs do not experience much acceleration in the HI FOV. For $\lambda=0^\circ$, 46\% of events have -1$<A<+$1\,m\,s$^{-2}$, for $\lambda=30^\circ$ the corresponding value is 48\% and for $\lambda=90^\circ$ it is 51\%. Although accelerations tend to be small, their distribution depends quite strongly on the chosen half width. For $0^\circ$ half-width, 115 (77\%) of CMEs are accelerating and 34 (23\%) are decelerating, for $30^\circ$ half-width 98 (66\%) of CMEs are accelerating and 51 (34\%) are decelerating and for $90^\circ$ half-width 79 (53\%) of CMEs are accelerating and 70 (47\%) are decelerating. That most CMEs are accelerating in the HI FOV is inconsistent with results established previously: whilst \citet{1999StCyr} show the majority of CMEs to be accelerating within 2.44R$_\odot$, \citet{2009Gopalswamy} show that almost all have stopped accelerating by 32R$_\odot$. The mean acceleration for CMEs analysed using $\lambda=0^\circ$ is 6\,m\,s$^{-2}$, which is skewed well away from zero by CMEs that possess unphysically large accelerations that result from fitting with small half-widths, as was discussed in the previous section. In panel c of Figure \ref{fig:ssse_cartoon}, for example, the location of the CME shown in red is derived using a half-width of $40^\circ$. A CME fitted with a half-width of $0^\circ$ will be further from the Sun than the apex of that 40$^\circ$ half-width CME, at the points where the dashed lines intersect, whilst the apex of a CME with a half-width greater than $40^\circ$ will be closer to the Sun. The SSSE method using $\lambda=0^\circ$ can result in increasingly large speeds and large accelerations, as is seen in panels c and d of Figure \ref{fig:ssse_kinematics}. This effect tends to be less apparent for larger half-widths; in the cases of $\lambda=30^\circ$ and $\lambda=90^\circ$, the mean accelerations are 1 and 0\,m\,s$^{-2}$, respectively. Regardless of the half-width chosen, we still find, as noted above, that the number of accelerating CMEs is always greater than the number of those decelerating. Even for an intermediate half-width of $30^\circ$, almost two thirds of the events appear to experience acceleration. The results do suggest that, in general, CMEs continue to experience acceleration within the HI FOVs, however, this is usually not significant in magnitude.

Panels b, d and f of Figure \ref{fig:deflection} show the distributions of CME deflections in ecliptic longitude, determined using SSSE analysis with respective half-widths of 0$^\circ$, 30$^\circ$ and 90$^\circ$. The deflection refers to the difference between the final and initial longitudinal position of the CME apex. For all three half-widths, significant deflections are often seen. For $\lambda=0^\circ$, 33\% of CMEs deflect by more than $\pm10^\circ$; corresponding values for $\lambda=30^\circ$ and $90^\circ$ are 45\% and 54\%, respectively. In some cases we appear to observe deflections of up to $90^\circ$, far in excess of the maximum deflection of $29^\circ$ observed by \citet{2014Isavnin} and indeed of the $20^\circ$ deflection by \citet{2014Wang}. This is in contradiction to any known physical process and is instead the result of inadequacies with the assumptions of the analysis method. As was discussed in the previous section, and is seen in Figure \ref{fig:ssse_kinematics}, the SSSE method is sometimes poor at determining the orientation of wider CMEs when they are close to the Sun, due to the fact that they expand at a rate greater than self-similarity \citep{2013Liu}. Indeed, because of this issue, it is difficult to ascertain how much of the longitudinal deflections determined using SSSE analysis are due to inaccuracies in the method, because we are unable to accurately measure their initial longitude. However, the propagation direction measurements become more constant as the CME propagates further into the heliosphere and so determining this value further out into the HI FOV is likely to provide a more reliable estimate of the ultimate CME propagation direction.
 
\begin{figure}
\centering
\includegraphics[width=\textwidth]{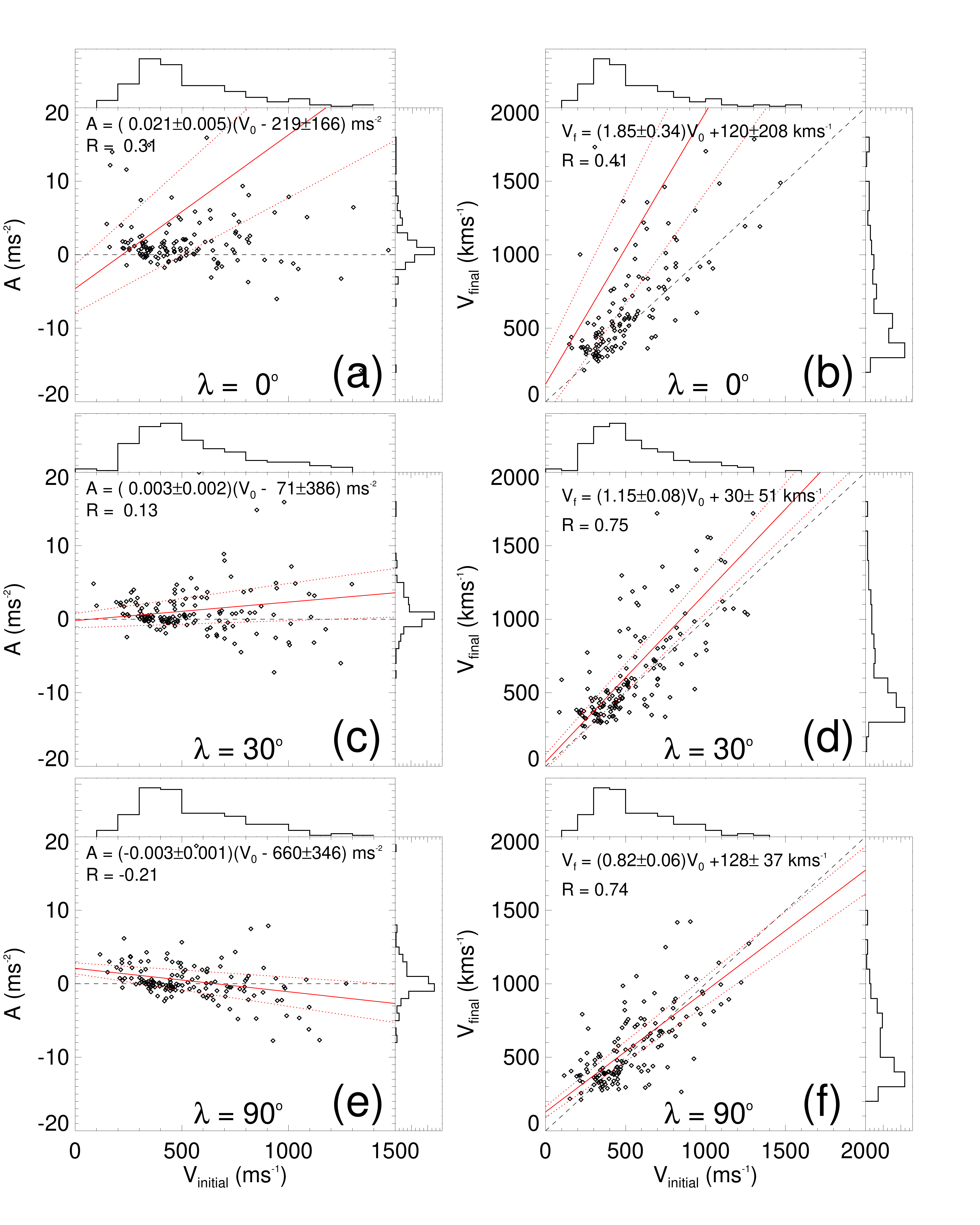}
\caption{Panels \textbf{a}, \textbf{c} and \textbf{e} show the relationship between CME initial velocity and CME acceleration, determined for respective SSSE half-widths of $\lambda=0^\circ$, $30^\circ $ and $90^\circ$. Panels \textbf{b}, \textbf{d} and \textbf{f} show the relationship between initial and final velocity, again for $\lambda=0^\circ$, $30^\circ $ and $90^\circ$. Over-plotted in each case is the regression line and printed are the correlation coefficient. On all plots the histograms on the top and right indicate the distributions of $x-$ and $y-$ parameter values.}
\label{fig:acceleration}
\end{figure}

\noindent
Figure \ref{fig:acceleration} shows a comparison between the initial CME velocities and the CME acceleration in panels a, c and e, using SSSE with respective half-widths of $\lambda=0^\circ$, 30$^\circ$ and $90^\circ$. Panel a shows no clear correlation between the initial velocity and the measured acceleration when using $\lambda=0^\circ$ to model the CME and the regression line is strongly skewed by CMEs with high acceleration values exceeding the plot range. Likewise, there is little correlation between initial velocity and acceleration for $\lambda=30^\circ$ in panel c. For $\lambda=90^\circ$, in panel e, there appears to be a slight tendency for the slowest CMEs to experience a positive acceleration and, conversely, for the faster CMEs to experience a deceleration, as expected \citep{2004Yashiro,2009Gopalswamy}. As noted previously, smaller half-widths, particularly $\lambda=0^\circ$, can lead to unphysically large accelerations. Panel e is consistent with the idea that CMEs experience drag from the ambient solar wind that causes their speed to tend towards the typical solar wind speed, which is not the case for panels a and c. This may suggest that a half-width between $30^\circ$ and $90^\circ$ is a better representation of the observed CMEs. Indeed, \citet{2004Yashiro} show that the mean width of CMEs observed using LASCO increases from 46$^\circ$ in 1996 (solar minimum) to 57$^\circ$ in 2000 (solar maximum), however this width is measured in PA and non longitude. The regression line in panel e suggests that the juncture between CMEs that accelerate and those that decelerate corresponds to an initial speed of $660\pm346$\,km\,s$^{-1}$, which, although rather imprecise, does correspond to the typical slow solar wind velocity of around 400\,--\,500\,km\,s$^{-1}$.

The right hand panels (b, d and f) of Figure \ref{fig:acceleration} show a comparison between the initial CME speed and the final CME speed derived using half-widths of $0^\circ$, $30^\circ$ and $90^\circ$, respectively. In the case of $\lambda=0^\circ$ (panel b) there is little correlation between initial and final velocities, which is due to the fact that many CMEs are found to have unphysically high final speeds when using this half-width, regardless of their initial speed. The regression line is strongly skewed by CMEs with final velocities exceeding 2000\,km\,s$^{-1}$. In fact, 22 CMEs (15\%) analysed assuming $\lambda=0^\circ$ have final velocities that exceed the 2000\,km\,s$^{-1}$ upper limit of the plot, whilst only three CMEs do so for $\lambda=30^\circ$ and only one for $\lambda=90^\circ$. For $\lambda=30^\circ$ (panel d) and $90^\circ$ (panel f), there is a strong correlation between initial and final velocity, with respective correlation coefficients of 0.75 and 0.74. The CME final velocity is less spread than that of initial velocity in each case, which can be seen in the histograms at the top and right of each panel. This can be explained by the idea that CMEs tend towards the ambient solar wind speed. For the CMEs analysed using $\lambda=90^\circ$, 60\% of slower events, those that have an initial velocity below 500\,km\,s$^{-1}$, are accelerating and 57\% of those with an initial velocity above this value are decelerating. In the case of $\lambda=30^\circ$, the majority of both slower and faster events are accelerating, which is inconsistent with established CME behaviour (e.g. \citealp{2004Yashiro,2009Gopalswamy}) and suggests that this may demonstrate inadequacies in the use of this geometry to describe the CMEs.

\subsection{A Comparison of Single-Spacecraft and Stereoscopic Techniques}

\begin{figure}
\centering 
\includegraphics[width=\textwidth]{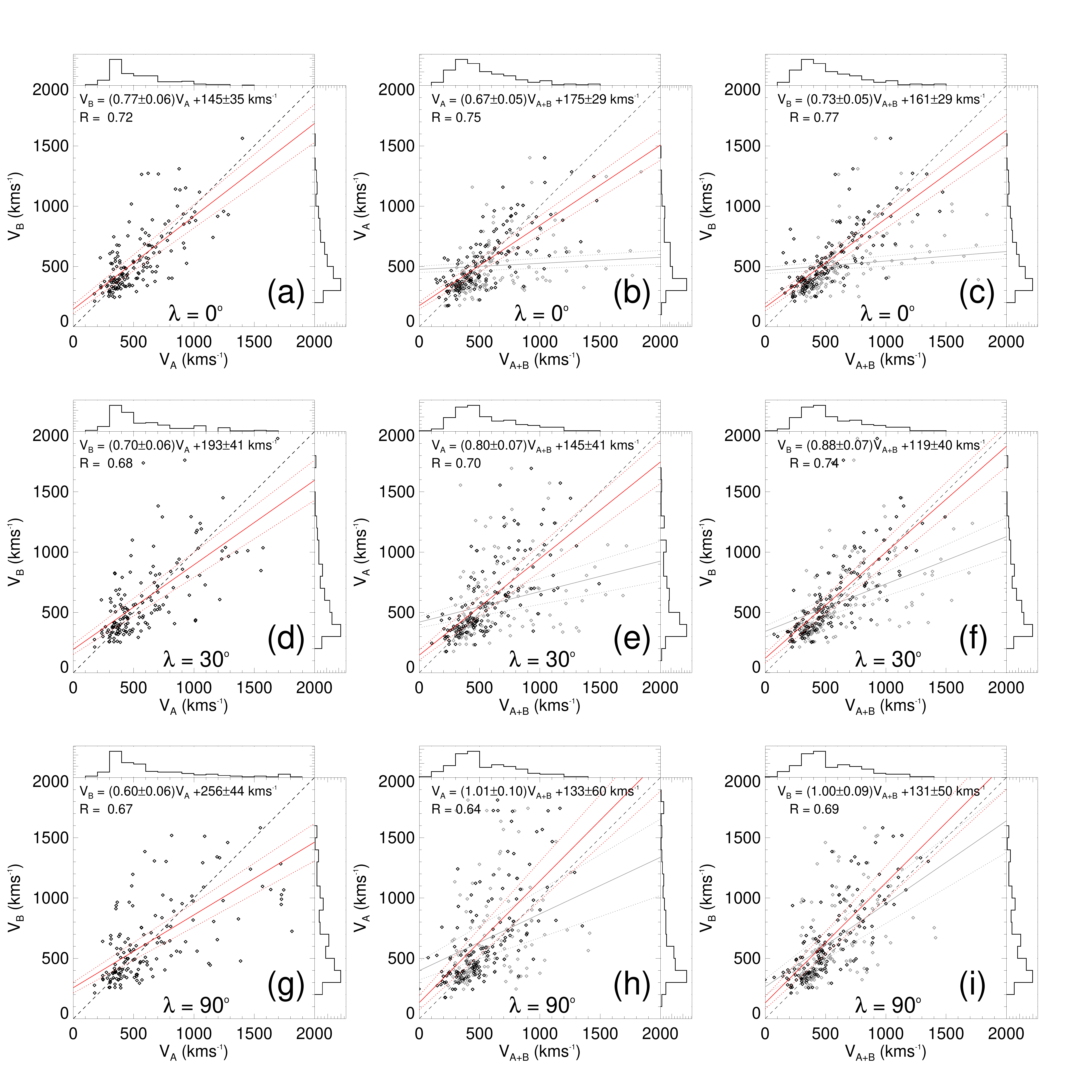}
\caption{Scatter plots showing a comparison between velocities determined from three methods for each of the 151 CMEs to which the stereoscopic analysis was applied. (\textbf{a}), (\textbf{b}) and (\textbf{c}) use $\lambda=0^\circ$, (\textbf{d}), (\textbf{e}) and (\textbf{f}) use $\lambda=30^\circ$ and (\textbf{g}), (\textbf{h}) and (\textbf{i}) use $\lambda=90^\circ$. Panels in the left-hand, centre and right-hand columns show $V_A$ versus $V_B$, $V_{A+B}$ versus $V_A$ and $V_{A+B}$ versus $V_B$, respectively. Where $V_A$ and $V_B$ are the speeds from single-spacecraft fits to HI-A and HI-B time--elongation profiles and $V_{A+B}$ is the initial velocity derived from stereoscopic analysis applied to the same events. A linear regression line is over-plotted in each figure, as is its equation. The histograms plotted on the top and right of each panel show the distribution of parameter values on each axis. Light grey points instead represent the CME final velocities derived using the stereoscopic analysis, plotted against the single-spacecraft speeds.}
\label{fig:speed_comparison}
\end{figure}

\noindent
Figure \ref{fig:speed_comparison} shows a comparison between the velocities determined for each CME from single-spacecraft, SSE, and stereoscopic, SSSE, analysis, resulting from three assumed geometries corresponding to $\lambda=0^\circ$, $30^\circ$ and $90^\circ$ (equivalent to the single--spacecraft FPF, SSEF30 and HMF techniques); $v_A$ and $v_B$ are the velocities resulting from the single-spacecraft fitting methods applied to STEREO-A and STEREO-B time--elongation profiles, respectively, and $v_{A+B}$ is the initial speed from the stereoscopic method. Panels a, b and c (top row) correspond to results using a half-width of 0$^\circ$, d, e and f (middle row) use 30$^\circ$ and g, h and i (bottom row) use 90$^\circ$. Each column presents panels corresponding to the three combinations of pairs of $v_A$, $v_B$ and $v_{A+B}$. The histograms at the top and left of each panel show the speed distribution corresponding to the $x$ and $y$ parameters plotted in that panel. The correlation coefficient, $R$, between each pair of velocity measurement ranges between 0.64 (panel h) and 0.77 (panel c), showing that there is reasonable agreement between speeds derived from all three methods, for each half-width. The best agreement is found for $\lambda=0^\circ$ and the worst agreement for $\lambda=90^\circ$. The CME final speeds derived using the stereoscopic analysis method, over plotted in light grey (panels in the second and third columns), are found to show a much poorer correlation with the single-spacecraft speeds (with $R$ ranging from 0.26 in panel b to 0.50 in panels f and i). This correlation is worst for $\lambda=0^\circ$ (panels b and c) and improves with increasing half-width; the best correlation is seen in panels h and i, using a $\lambda=90^\circ$. As shown in Figure \ref{fig:ssse_kinematics}, for example, the final speeds derived using the stereoscopic method with $\lambda=0^\circ$ are often unphysically high. However, even for $\lambda=90^\circ$, the correlation between the final velocity derived from stereoscopic analysis and from single-spacecraft fitting is still far worse than that of the initial velocities. In fact the correlation coefficient has a value of only 0.26 between $v_{A+B}$ (final) and $V_A$ and a value of 0.38 between $v_{A+B}$ (final) and $v_B$. This is consistent with the results of \citet{2013Liu}, who find an "apparent late acceleration" for CMEs fitted using FPF (equivalent to SSEF with $\lambda=0^\circ$). They show that the HMF ($\lambda=90^\circ$) method can reduce this effect, but, however, that it can still produce an overestimate of CME speed further out into the heliosphere. Single-spacecraft derived speeds of those CMEs included in the \textsf{HIGeoCAT} catalogue that impact spacecraft throughout the inner heliosphere were compared to in-situ signatures by \cite{2017Mostl}. For those predicted impacts of \textsf{HIGeoCAT} CMEs that matched with in-situ impacts, the predicted arrival times (derived using SSE with $\lambda=30^\circ$) were found to be 2.4$\pm$17.1\,h early for HI-A and 2.7$\pm$16.0\,h early for HI-B, for events within a time window of $\pm$1\,day. The \textsf{HiGeoCAT} speeds were on average 191$\pm$341\,km\,s$^{-1}$ greater than those measured in-situ for HI-A CMEs and 245$\pm$446\,km\,s$^{-1}$ greater for HI-B CMEs. However, a similar study has not been perform using the CMEs in the \textsf{HIJoinCAT}, which are analysed using the SSSE method, and which would provide a measure of ground truth with which to compare all three fitting methods.

\begin{figure}
\centering
\includegraphics[width=\textwidth]{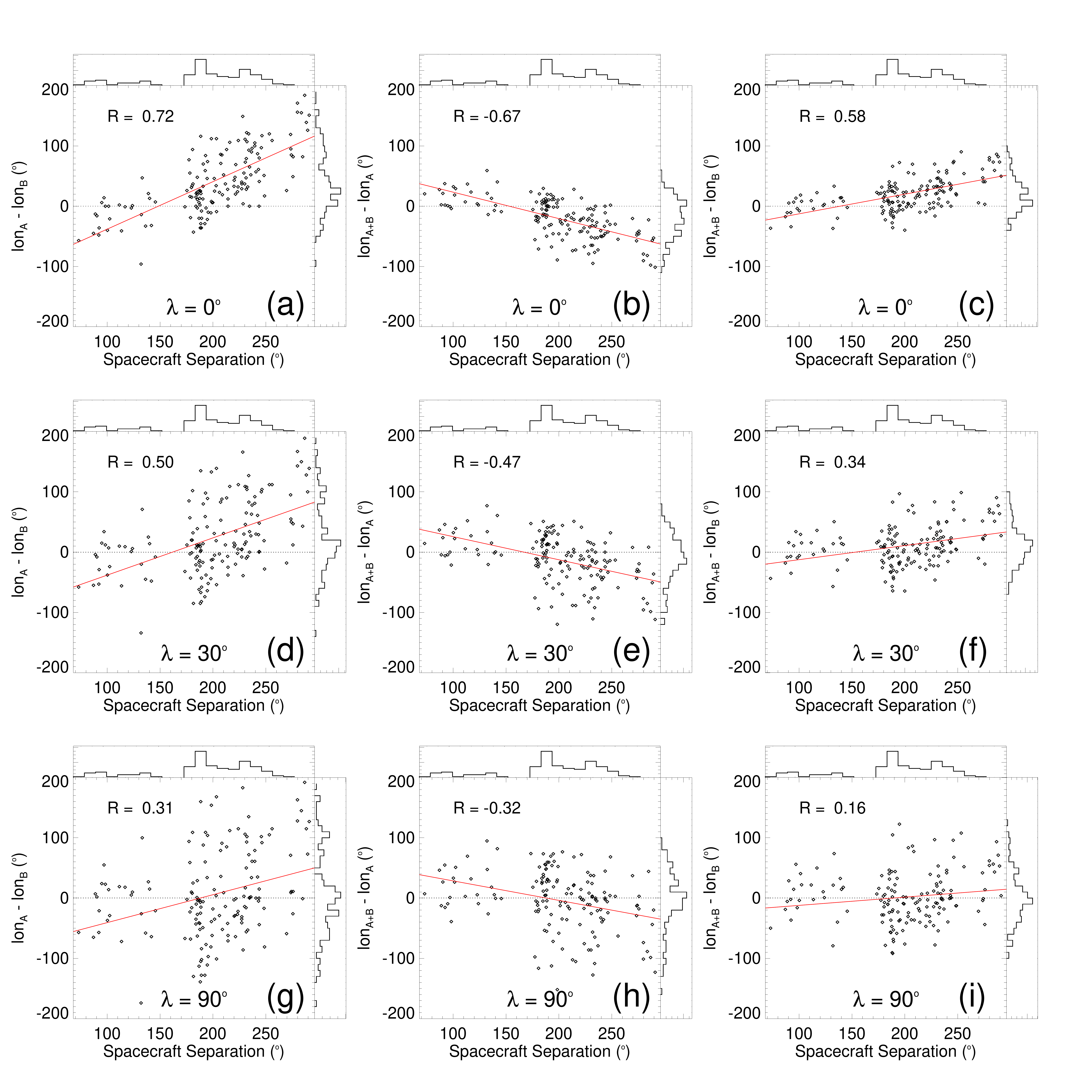}
\caption{Scatter-plots showing the difference in CME apex longitude in HEEQ coordinates between each combination of the three methods versus the spacecraft separation angle. (\textbf{a}), (\textbf{b}) and (\textbf{c})  use $\lambda=0^\circ$, (\textbf{d}), (\textbf{e}) and (\textbf{f}) use $\lambda=30^\circ$ and (\textbf{g}), (\textbf{h}) and (\textbf{i}) use $\lambda=90^\circ$. lon$_A$ and lon$_B$ are the CME apex longitude from the methods using SSEF with STEREO-A and STEREO-B and lon$_{A+B}$ is the apex longitude from SSSE analysis with both STEREO spacecraft. In each plot the regression line is over-plotted and the histograms represent the distribution of data points on each axis.}
\label{fig:phi_comparison}
\end{figure}

Figure \ref{fig:phi_comparison} shows a comparison between the longitudinal propagation angles determined for each CME, from each of the three methods. Here, lon$_A$ and lon$_B$ are the derived longitude of the CME apex in Heliocentric Earth Equatorial (HEEQ) coordinates from single-spacecraft analysis, whilst lon$_{A+B}$ is the final CME apex longitude derived from stereoscopic analysis in the same coordinate system. The top, middle and bottom row of panels correspond to a half-width of 0$^\circ$, $30^\circ$ and $90^\circ$ respectively. Each plot shows the difference in longitude between two of the three fitting methods, as a function of spacecraft separation angle. The histogram at the top of each plot shows the distribution of CMEs as a function of spacecraft separation angle, which increases as a function of time, and therefore correspond approximately to the dark grey distribution in Figure \ref{fig:stereo_hist}. The total range of separation angles over which the CMEs are observed is 73$^\circ$ to 291$^\circ$ and the majority of CMEs (78\%) occur once the spacecraft separation is greater than 180$^\circ$ because this equates to solar maximum. The longitudes derived from single-spacecraft analysis are in fairly poor agreement with each other; moreover, neither agree well with the results from stereoscopic analysis. In addition, the discrepancy between each set of results appears to show a systematic variation with spacecraft separation. For each row (i.e. different half-width) the difference between the SSE$_A$ and SSE$_B$ longitudes (left-hand column) suggests that the methods produce a bias towards a certain range of CME propagation directions relative to the spacecraft. As the spacecraft move apart in longitude, in opposite directions, this bias results in the observed correlation between separation angle and longitude difference. In Sections 3.2 and 3.3 of Article 2, we studied the distribution of CME propagation angle, $\phi$, relative to the spacecraft, for all 1455 CMEs in version 5 of the single-spacecraft fitting catalogue \textsf{HiGeoCAT}. We showed that of each distribution was peaked at around 78$^\circ$ ($\lambda=0^\circ$), 72$^\circ$ ($\lambda=30^\circ$) and 84$^\circ$ ($\lambda=90^\circ$) for STEREO-A CMEs and 72$^\circ$ ($\lambda=0^\circ$), 69$^\circ$ ($\lambda=30^\circ$) and 77$^\circ$ ($\lambda=90^\circ$) for STEREO-B CMEs. This is believed to be due to two effects. The first is an observational effect, whereby it is somewhat easier to observe CMEs travelling close to the Thomson surface \citep{2009aTappin}. The second is an inherent bias in the single-spacecraft fitting models, found by \citet{2010bLugaz}, who show that assuming $\lambda=0^\circ$ for a wide CME (with $90^\circ$) causes a bias towards propagation directions close to $60^\circ$ from the Sun-spacecraft line, for CMEs that propagate at more than $\pm20^\circ$ from this direction. Systematic effects are also seen in the second (panels b, e and h) and third (panels c, f and i) columns, where we compare the stereoscopic results to each of the single-spacecraft results. However, they are less significant because the stereoscopic method only suffers from the Thomson surface effect and not the bias from the single-spacecraft model assumptions. The regression line in panel a of Figure \ref{fig:phi_comparison} crosses zero on the y-axis when the spacecraft separation is close to $150^\circ$, which is the sum of the median propagation directions found from SSEF with $\lambda=0^\circ$ (FPF) for each spacecraft in Article 2. That is, these biases cause the FPF longitudes to coincide when the spacecraft separation is $150^\circ$.

\section{Summary}
From all CMEs observed by HI on STEREO-A and STEREO-B, whilst contact still existed with the latter, we identify 273 CMEs, occurring between 31 August 2008 and 02 April 2014, that are observed by both. We apply SSSE analysis techniques to these CMEs in order to determine their kinematic behaviour. During this time the spacecraft longitude separation increased from 73$^\circ$ to 291$^\circ$. This period spans approximately half a solar cycle, beginning at solar minimum and ending at the peak of Solar Cycle 24.

The main conclusions are summarised as follows:

\begin{enumerate}[i.]
\item The SSSE method fails when the CME passes between the observing spacecraft and the lines-of-sight of the CME leading edge are close to parallel because small errors in elongation translate to large errors in determining CME position. We therefore apply the technique to just 151 CMEs, 78\% of which occur close to solar maximum, after the spacecraft separation exceeds 180$^\circ$. These data are too few to perform a thorough investigation of the optimal spacecraft configuration with which to analyse CMEs using the SSSE method. However, the results show that two spacecraft situated at L4 and L5, with a separation of $120^\circ$, would be a feasible configuration to track an Earth-directed CME, until its front reaches $60^\circ$ elongation, where the LOSs would become parallel.

\item Accelerations derived using the SSSE technique are much higher when the smaller half-width is chosen. For $\lambda=0^\circ$, 76\% of CMEs were found to have positive acceleration and 15\% showed a final velocity exceeding 2000\,km\,s$^{-1}$. For $\lambda=30^\circ$, 66\% of all CMEs were found to be accelerating, regardless of their initial velocity. Conversely, using $90^\circ$ results in approximately half of CMEs accelerating and half decelerating (52\% versus 48\%, respectively), suggesting that this model agrees best with the average CME width of 47\,--\,60$^\circ$ \citep{2004Yashiro}. For slower CMEs, with initial speeds below 500\,km\,s$^{-1}$, 60\% are seen to accelerate and 57\% of CMEs faster than 500\,km\,s$^{-1}$ are seen to decelerate when using $\lambda=90^\circ$. This is consistent with the well-established understanding that drag between CMEs and the background solar wind causes the CME speed to tend towards the ambient solar wind speed. These results suggest that CMEs may undergo acceleration well into the HI FOV, contrary to the main assumption used in SSEF analysis.

\item The longitude of CMEs is found to vary greatly when they are close to the Sun, due to the fact that our SSSE analysis does not account for CME over-expansion. As the CME is tracked to larger elongation angles, the propagation direction is found to approach a constant value. It is therefore difficult to draw meaningful information about CME deflections, because we cannot accurately know their initial longitude. However, the final longitude for $\lambda=90^\circ$ is expected to provide a good estimate of the ultimate CME propagation direction.

\item The velocity for each of the 151 CMEs is determined using three different means: SSEF using STEREO-A data, SSEF using STEREO-B data and SSSE analysis using data from both spacecraft. Each technique is applied using three different half-widths to fit the CMEs: 0$^\circ$, 30$^\circ$ and 90$^\circ$. Agreement between initial CME speed is good between all methods, however the final CME speed derived from SSSE analysis does not agree with that from SSEF. This is in part due to the over-estimation of CME acceleration when using a small half-width in the SSSE analysis.

\item  Similarly, we compare the difference in HEEQ longitude of the CME apex between each pair of fitting methods, again using $\lambda=0^\circ$, $30^\circ$ and $90^\circ$. The agreement between the SSEF methods from each spacecraft is poor, with differences close to 180$^\circ$ in the worst cases. The effect is systematic and is a function of spacecraft separation angle, which is due to three causes. Firstly, projection effects caused by Thomson scattering; secondly, biases in the CME direction determined when using a small half-width, as identified by \citet{2010bLugaz}; and, thirdly, the incorrect assumptions employed by the single-spacecraft fitting methods.

\end{enumerate}

White-light heliospheric imaging offers a unique way to track CMEs through the inner heliosphere. In Article 1 we presented a catalogue of interplanetary CMEs that, at the time of writing, contains over 2000 events and spans an entire solar cycle. Many of these CMEs have been studied using single-spacecraft analysis techniques in Article 2 and 151, presented here, have been analysed using stereoscopic observations. If we wish to track CMEs in the heliosphere, for the purposes of both science and space weather, there are, however, many limitations that result from doing so with observations from just one spacecraft, and many still with observations from two. Many of the limitations identified in this article are possible to address by modifying the way that the SSSE analysis is applied to the data. For example, modelling a CME with super self-similar expansion would account for CME over-expansion in the early propagation phase. Alternatively, it may be preferable to analyse the CME in coronagraphs separately using models, such as GCS, that can measure CME expansion, before moving to SSSE analysis at greater distances from the Sun. With the upcoming launches of the PUNCH mission in Earth orbit, ESA's Lagrange mission to L5, as well as the continued coverage from STEREO-A and the recent launches of Parker Solar Probe and Solar Orbiter we are entering an era of unprecedented coverage from wide-angle imagers. It will therefore soon be possible to scrutinise these methods further: for example, the extra information available from three or more vantage points will allow the measurement of non-circular CME fronts and will greatly limit the cases where the SSSE method fails due to parallel LOS observations or when observing CMEs at small elongation angles close to the Sun. Further to this, Solar Orbiter and Parker Solar Probe will observe from higher latitues giving a truly three-dimensional view of CMEs when combined with observations from the ecliptic. The PUNCH mission possesses the unprecedented advantage of measuring polarisation of Thomson-scattered light in heliospheric observations, which provides a further means to constrain the location of observed features along the LOS (e.g. \citet{2016aDeForest}).

\begin{acknowledgements}
We acknowledge support from the European Union FP7--SPACE--2013--1 programme for the HELCATS project (\#606692). The HI instruments on STEREO were developed by a consortium that comprised the Rutherford Appleton Laboratory (UK), the University of Birmingham (UK), Centre Spatial de Li\`ege (CSL, Belgium) and the Naval Research Laboratory (NRL, USA). The STEREO/SECCHI project, of which HI is a part, is an international consortium led by NRL. We recognise the support of the UK Space Agency for funding STEREO/HI operations in the UK. C.M. thanks the Austrian Science Fund (FWF): P31521-N27, P31659-N27. 
\end{acknowledgements}

\noindent
{\footnotesize\textbf{Disclosure of Potential Conflicts of Interest} The authors declare that they have no conflicts of interest.}

\bibliographystyle{spr-mp-sola}
\bibliography{biblio}

\begin{thebibliography}{76}
\ifx\bisbn     \undefined \def\bisbn  #1{ISBN #1}\fi
\ifx\binits    \undefined \def\binits#1{#1}\fi
\ifx\bauthor   \undefined \def\bauthor#1{#1}\fi
\ifx\batitle   \undefined \def\batitle#1{#1}\fi
\ifx\bjtitle   \undefined \def\bjtitle#1{\textit{#1}}\fi
\ifx\bvolume   \undefined \def\bvolume#1{\textbf{#1}}\fi
\ifx\byear     \undefined \def\byear#1{#1}\fi
\ifx\bissue    \undefined \def\bissue#1{#1}\fi
\ifx\bfpage    \undefined \def\bfpage#1{#1}\fi
\ifx\blpage    \undefined \def\blpage #1{#1}\fi
\ifx\burl      \undefined \def\burl#1{\textsf{#1}}\fi
\ifx\href      \undefined \def\href#1#2{\textsf{#2}}\fi
\ifx\betal     \undefined \def\betal{\textit{et al.}}\fi
\ifx\bctitle   \undefined \def\bctitle#1{#1}\fi
\ifx\beditor   \undefined \def\beditor#1{#1}\fi
\ifx\bbtitle   \undefined \def\bbtitle#1{\textit{#1}}\fi
\ifx\bedition  \undefined \def\bedition#1{#1}\fi
\ifx\bseriesno \undefined \def\bseriesno#1{\textbf{#1}}\fi
\ifx\blocation \undefined \def\blocation#1{#1}\fi
\ifx\bsertitle \undefined \def\bsertitle#1{\textit{#1}}\fi
\ifx\bsnm      \undefined \def\bsnm#1{#1}\fi
\ifx\bsuffix   \undefined \def\bsuffix#1{#1}\fi
\ifx\bparticle \undefined \def\bparticle#1{#1}\fi
\ifx\barticle  \undefined \def\barticle#1{}\fi
\ifx\binstitute  \undefined \def\binstitute#1{#1}\fi
\ifx\bpublisher  \undefined \def\bpublisher#1{#1}\fi
\ifx\doiurl    \undefined
  \def\doiurl#1{\href{http://dx.doi.org/#1}{\textsf{DOI}}}\fi
\ifx\arxivurl  \undefined
  \def\arxivurl#1{\href{http://arxiv.org/abs/#1}{\textsf{arXiv}}}\fi
\ifx\adsurl    \undefined
  \def\adsurl#1{\href{http://adsabs.harvard.edu/abs/#1}{\textsf{ADS}}}\fi
\ifx\botherref \undefined \def\botherref#1{}\fi
\ifx\url       \undefined \def\url#1{\textsf{#1}}\fi
\ifx\bchapter  \undefined \def\bchapter#1{}\fi
\ifx\bbook     \undefined \def\bbook#1{}\fi
\ifx\bcomment  \undefined \def\bcomment#1{#1}\fi
\ifx\oauthor   \undefined \def\oauthor#1{#1}\fi
\ifx\citeauthoryear \undefined\def \citeauthoryear#1{#1}\fi
\ifx\endbibitem\undefined \def\endbibitem{}\fi
\ifx\bconflocation  \undefined \def\bconflocation#1{#1} \fi

\bibitem[\protect\citeauthoryear{Barnes \textit{et~al.}}{2019}]{2019Barnes}
\begin{barticle}
\bauthor{\bsnm{Barnes}, \binits{D.}},
\bauthor{\bsnm{Davies}, \binits{J.A.}},
\bauthor{\bsnm{Harrison}, \binits{R.A.}},
\bauthor{\bsnm{Byrne}, \binits{J.P.}},
\bauthor{\bsnm{Perry}, \binits{C.H.}},
\bauthor{\bsnm{Bothmer}, \binits{V.}},
\bauthor{\bsnm{Eastwood}, \binits{J.P.}},
\bauthor{\bsnm{Gallagher}, \binits{P.T.}},
\bauthor{\bsnm{Kilpua}, \binits{E.K.J.}},
\bauthor{\bsnm{M{\"o}stl}, \binits{C.}},
\bauthor{\bsnm{Rodriguez}, \binits{L.}},
\bauthor{\bsnm{Rouillard}, \binits{A.P.}},
\bauthor{\bsnm{Odstr{\v{c}}il}, \binits{D.}}:
\byear{2019},
\batitle{{CMEs in the Heliosphere: II. A Statistical Analysis of the Kinematic
  Properties Derived from Single-Spacecraft Geometrical Modelling Techniques
  Applied to CMEs Detected in the Heliosphere from 2007 to 2017 by
  STEREO/HI-1}}.
\bjtitle{Solar Physics}
\bvolume{294}(\bissue{5}),
\bfpage{57}.
\doiurl{10.1007/s11207-019-1444-4}.
\end{barticle}
\endbibitem

\bibitem[\protect\citeauthoryear{Bothmer and Schwenn}{1997}]{1997Bothmer}
\begin{barticle}
\bauthor{\bsnm{Bothmer}, \binits{V.}},
\bauthor{\bsnm{Schwenn}, \binits{R.}}:
\byear{1997},
\batitle{The structure and origin of magnetic clouds in the solar wind}.
\bjtitle{Annales Geophysicae}
\bvolume{16}(\bissue{1}),
\bfpage{1}.
\doiurl{10.1007/PL00021390}.
\burl{https://doi.org/10.1007/PL00021390}.
\end{barticle}
\endbibitem

\bibitem[\protect\citeauthoryear{{Brueckner}
  \textit{et~al.}}{1995}]{1995Brueckner}
\begin{barticle}
\bauthor{\bsnm{{Brueckner}}, \binits{G.E.}},
\bauthor{\bsnm{{Howard}}, \binits{R.A.}},
\bauthor{\bsnm{{Koomen}}, \binits{M.J.}},
\bauthor{\bsnm{{Korendyke}}, \binits{C.M.}},
\bauthor{\bsnm{{Michels}}, \binits{D.J.}},
\bauthor{\bsnm{{Moses}}, \binits{J.D.}},
\bauthor{\bsnm{{Socker}}, \binits{D.G.}},
\bauthor{\bsnm{{Dere}}, \binits{K.P.}},
\bauthor{\bsnm{{Lamy}}, \binits{P.L.}},
\bauthor{\bsnm{{Llebaria}}, \binits{A.}},
\bauthor{\bsnm{{Bout}}, \binits{M.V.}},
\bauthor{\bsnm{{Schwenn}}, \binits{R.}},
\bauthor{\bsnm{{Simnett}}, \binits{G.M.}},
\bauthor{\bsnm{{Bedford}}, \binits{D.K.}},
\bauthor{\bsnm{{Eyles}}, \binits{C.J.}}:
\byear{1995},
\batitle{{The Large Angle Spectroscopic Coronagraph (LASCO)}}.
\bjtitle{\solphys}
\bvolume{162},
\bfpage{357}.
\doiurl{10.1007/BF00733434}.
\adsurl{1995SoPh..162..357B}.
\end{barticle}
\endbibitem

\bibitem[\protect\citeauthoryear{{Byrne} \textit{et~al.}}{2010}]{2010Byrne}
\begin{barticle}
\bauthor{\bsnm{{Byrne}}, \binits{J.P.}},
\bauthor{\bsnm{{Maloney}}, \binits{S.A.}},
\bauthor{\bsnm{{McAteer}}, \binits{R.T.J.}},
\bauthor{\bsnm{{Refojo}}, \binits{J.M.}},
\bauthor{\bsnm{{Gallagher}}, \binits{P.T.}}:
\byear{2010},
\batitle{{Propagation of an Earth-directed coronal mass ejection in three
  dimensions}}.
\bjtitle{Nature Communications}
\bvolume{1},
\bfpage{74}.
\doiurl{10.1038/ncomms1077}.
\adsurl{2010NatCo...1E..74B}.
\end{barticle}
\endbibitem

\bibitem[\protect\citeauthoryear{{Cremades, H.} and {Bothmer,
  V.}}{2004}]{2004Cremades}
\begin{barticle}
\bauthor{\bsnm{{Cremades, H.}}},
\bauthor{\bsnm{{Bothmer, V.}}}:
\byear{2004},
\batitle{On the three-dimensional configuration of coronal mass ejections}.
\bjtitle{A\&A}
\bvolume{422}(\bissue{1}),
\bfpage{307}.
\doiurl{10.1051/0004-6361:20035776}.
\burl{https://doi.org/10.1051/0004-6361:20035776}.
\end{barticle}
\endbibitem

\bibitem[\protect\citeauthoryear{{Davies} \textit{et~al.}}{2009}]{2009Davies}
\begin{barticle}
\bauthor{\bsnm{{Davies}}, \binits{J.A.}},
\bauthor{\bsnm{{Harrison}}, \binits{R.A.}},
\bauthor{\bsnm{{Rouillard}}, \binits{A.P.}},
\bauthor{\bsnm{{Sheeley}}, \binits{N.R.}},
\bauthor{\bsnm{{Perry}}, \binits{C.H.}},
\bauthor{\bsnm{{Bewsher}}, \binits{D.}},
\bauthor{\bsnm{{Davis}}, \binits{C.J.}},
\bauthor{\bsnm{{Eyles}}, \binits{C.J.}},
\bauthor{\bsnm{{Crothers}}, \binits{S.R.}},
\bauthor{\bsnm{{Brown}}, \binits{D.S.}}:
\byear{2009},
\batitle{{A synoptic view of solar transient evolution in the inner heliosphere
  using the Heliospheric Imagers on STEREO}}.
\bjtitle{\grl}
\bvolume{36},
\bfpage{L02102}.
\doiurl{10.1029}.
\adsurl{2009GeoRL..3602102D}.
\end{barticle}
\endbibitem

\bibitem[\protect\citeauthoryear{{Davies} \textit{et~al.}}{2012}]{2012Davies}
\begin{barticle}
\bauthor{\bsnm{{Davies}}, \binits{J.A.}},
\bauthor{\bsnm{{Harrison}}, \binits{R.A.}},
\bauthor{\bsnm{{Perry}}, \binits{C.H.}},
\bauthor{\bsnm{{M{\"o}stl}}, \binits{C.}},
\bauthor{\bsnm{{Lugaz}}, \binits{N.}},
\bauthor{\bsnm{{Rollett}}, \binits{T.}},
\bauthor{\bsnm{{Davis}}, \binits{C.J.}},
\bauthor{\bsnm{{Crothers}}, \binits{S.R.}},
\bauthor{\bsnm{{Temmer}}, \binits{M.}},
\bauthor{\bsnm{{Eyles}}, \binits{C.J.}},
\bauthor{\bsnm{{Savani}}, \binits{N.P.}}:
\byear{2012},
\batitle{{A Self-similar Expansion Model for Use in Solar Wind Transient
  Propagation Studies}}.
\bjtitle{\apj}
\bvolume{750},
\bfpage{23}.
\doiurl{10.1088/0004-637X/750/1/23}.
\adsurl{2012ApJ...750...23D}.
\end{barticle}
\endbibitem

\bibitem[\protect\citeauthoryear{{Davies} \textit{et~al.}}{2013}]{2013Davies}
\begin{barticle}
\bauthor{\bsnm{{Davies}}, \binits{J.A.}},
\bauthor{\bsnm{{Perry}}, \binits{C.H.}},
\bauthor{\bsnm{{Trines}}, \binits{R.M.G.M.}},
\bauthor{\bsnm{{Harrison}}, \binits{R.A.}},
\bauthor{\bsnm{{Lugaz}}, \binits{N.}},
\bauthor{\bsnm{{M{\"o}stl}}, \binits{C.}},
\bauthor{\bsnm{{Liu}}, \binits{Y.D.}},
\bauthor{\bsnm{{Steed}}, \binits{K.}}:
\byear{2013},
\batitle{{Establishing a Stereoscopic Technique for Determining the Kinematic
  Properties of Solar Wind Transients based on a Generalized Self-similarly
  Expanding Circular Geometry}}.
\bjtitle{\apj}
\bvolume{777},
\bfpage{167}.
\doiurl{10.1088/0004-637X/777/2/167}.
\adsurl{2013ApJ...777..167D}.
\end{barticle}
\endbibitem

\bibitem[\protect\citeauthoryear{{Davis}, {Kennedy}, and
  {Davies}}{2010}]{2010Davis}
\begin{barticle}
\bauthor{\bsnm{{Davis}}, \binits{C.J.}},
\bauthor{\bsnm{{Kennedy}}, \binits{J.}},
\bauthor{\bsnm{{Davies}}, \binits{J.A.}}:
\byear{2010},
\batitle{{Assessing the Accuracy of CME Speed and Trajectory Estimates from
  STEREO Observations Through a Comparison of Independent Methods}}.
\bjtitle{\solphys}
\bvolume{263},
\bfpage{209}.
\doiurl{10.1007/s11207-010-9535-2}.
\adsurl{2010SoPh..263..209D}.
\end{barticle}
\endbibitem

\bibitem[\protect\citeauthoryear{{DeForest}
  \textit{et~al.}}{2016}]{2016aDeForest}
\begin{barticle}
\bauthor{\bsnm{{DeForest}}, \binits{C.E.}},
\bauthor{\bsnm{{Howard}}, \binits{T.A.}},
\bauthor{\bsnm{{Webb}}, \binits{D.F.}},
\bauthor{\bsnm{{Davies}}, \binits{J.A.}}:
\byear{2016},
\batitle{{The utility of polarized heliospheric imaging for space weather
  monitoring}}.
\bjtitle{Space Weather}
\bvolume{14},
\bfpage{32}.
\doiurl{10.1002/2015SW001286}.
\adsurl{2016SpWea..14...32D}.
\end{barticle}
\endbibitem

\bibitem[\protect\citeauthoryear{{Eyles} \textit{et~al.}}{2003}]{2003Eyles}
\begin{barticle}
\bauthor{\bsnm{{Eyles}}, \binits{C.J.}},
\bauthor{\bsnm{{Simnett}}, \binits{G.M.}},
\bauthor{\bsnm{{Cooke}}, \binits{M.P.}},
\bauthor{\bsnm{{Jackson}}, \binits{B.V.}},
\bauthor{\bsnm{{Buffington}}, \binits{A.}},
\bauthor{\bsnm{{Hick}}, \binits{P.P.}},
\bauthor{\bsnm{{Waltham}}, \binits{N.R.}},
\bauthor{\bsnm{{King}}, \binits{J.M.}},
\bauthor{\bsnm{{Anderson}}, \binits{P.A.}},
\bauthor{\bsnm{{Holladay}}, \binits{P.E.}}:
\byear{2003},
\batitle{{The Solar Mass Ejection Imager (Smei)}}.
\bjtitle{\solphys}
\bvolume{217},
\bfpage{319}.
\doiurl{10.1023/B:SOLA.0000006903.75671.49}.
\adsurl{2003SoPh..217..319E}.
\end{barticle}
\endbibitem

\bibitem[\protect\citeauthoryear{{Eyles} \textit{et~al.}}{2009}]{2009Eyles}
\begin{barticle}
\bauthor{\bsnm{{Eyles}}, \binits{C.J.}},
\bauthor{\bsnm{{Harrison}}, \binits{R.A.}},
\bauthor{\bsnm{{Davis}}, \binits{C.J.}},
\bauthor{\bsnm{{Waltham}}, \binits{N.R.}},
\bauthor{\bsnm{{Shaughnessy}}, \binits{B.M.}},
\bauthor{\bsnm{{Mapson-Menard}}, \binits{H.C.A.}},
\bauthor{\bsnm{{Bewsher}}, \binits{D.}},
\bauthor{\bsnm{{Crothers}}, \binits{S.R.}},
\bauthor{\bsnm{{Davies}}, \binits{J.A.}},
\bauthor{\bsnm{{Simnett}}, \binits{G.M.}},
\bauthor{\bsnm{{Howard}}, \binits{R.A.}},
\bauthor{\bsnm{{Moses}}, \binits{J.D.}},
\bauthor{\bsnm{{Newmark}}, \binits{J.S.}},
\bauthor{\bsnm{{Socker}}, \binits{D.G.}},
\bauthor{\bsnm{{Halain}}, \binits{J.-P.}},
\bauthor{\bsnm{{Defise}}, \binits{J.-M.}},
\bauthor{\bsnm{{Mazy}}, \binits{E.}},
\bauthor{\bsnm{{Rochus}}, \binits{P.}}:
\byear{2009},
\batitle{{The Heliospheric Imagers Onboard the STEREO Mission}}.
\bjtitle{\solphys}
\bvolume{254},
\bfpage{387}.
\doiurl{10.1007/s11207-008-9299-0}.
\adsurl{2009SoPh..254..387E}.
\end{barticle}
\endbibitem

\bibitem[\protect\citeauthoryear{{Fisher} \textit{et~al.}}{1981}]{1981Fisher}
\begin{barticle}
\bauthor{\bsnm{{Fisher}}, \binits{R.R.}},
\bauthor{\bsnm{{Lee}}, \binits{R.H.}},
\bauthor{\bsnm{{MacQueen}}, \binits{R.M.}},
\bauthor{\bsnm{{Poland}}, \binits{A.I.}}:
\byear{1981},
\batitle{{New Mauna Loa coronagraph systems}}.
\bjtitle{Applied Optics}
\bvolume{20},
\bfpage{1094}.
\doiurl{10.1364/AO.20.001094}.
\adsurl{1981ApOpt..20.1094F}.
\end{barticle}
\endbibitem

\bibitem[\protect\citeauthoryear{{Gopalswamy}}{2010}]{2010Gopalswamy}
\begin{bchapter}
\bauthor{\bsnm{{Gopalswamy}}, \binits{N.}}:
\byear{2010},
\bctitle{{Corona Mass Ejections: a Summary of Recent Results}}.
In: \beditor{\bsnm{{Dorotovic}}, \binits{I.}} (ed.)
\bbtitle{20th National Solar Physics Meeting},
\bfpage{108}.
\adsurl{2010nspm.conf..108G}.
\end{bchapter}
\endbibitem

\bibitem[\protect\citeauthoryear{Gopalswamy
  \textit{et~al.}}{2009}]{2009Gopalswamy}
\begin{barticle}
\bauthor{\bsnm{Gopalswamy}, \binits{N.}},
\bauthor{\bsnm{Yashiro}, \binits{S.}},
\bauthor{\bsnm{Michalek}, \binits{G.}},
\bauthor{\bsnm{Stenborg}, \binits{G.}},
\bauthor{\bsnm{Vourlidas}, \binits{A.}},
\bauthor{\bsnm{Freeland}, \binits{S.}},
\bauthor{\bsnm{Howard}, \binits{R.}}:
\byear{2009},
\batitle{The soho/lasco cme catalog}.
\bjtitle{Earth, Moon, and Planets}
\bvolume{104}(\bissue{1}),
\bfpage{295}.
\end{barticle}
\endbibitem

\bibitem[\protect\citeauthoryear{Gosling \textit{et~al.}}{1991}]{1991Gosling}
\begin{barticle}
\bauthor{\bsnm{Gosling}, \binits{J.T.}},
\bauthor{\bsnm{McComas}, \binits{D.J.}},
\bauthor{\bsnm{Phillips}, \binits{J.L.}},
\bauthor{\bsnm{Bame}, \binits{S.J.}}:
\byear{1991},
\batitle{Geomagnetic activity associated with earth passage of interplanetary
  shock disturbances and coronal mass ejections}.
\bjtitle{\jgr}
\bvolume{96}(\bissue{A5}),
\bfpage{7831}.
\doiurl{10.1029/91JA00316}.
\end{barticle}
\endbibitem

\bibitem[\protect\citeauthoryear{{Harrison}
  \textit{et~al.}}{2012}]{2012Harrison}
\begin{barticle}
\bauthor{\bsnm{{Harrison}}, \binits{R.A.}},
\bauthor{\bsnm{{Davies}}, \binits{J.A.}},
\bauthor{\bsnm{{M{\"o}stl}}, \binits{C.}},
\bauthor{\bsnm{{Liu}}, \binits{Y.}},
\bauthor{\bsnm{{Temmer}}, \binits{M.}},
\bauthor{\bsnm{{Bisi}}, \binits{M.M.}},
\bauthor{\bsnm{{Eastwood}}, \binits{J.P.}},
\bauthor{\bsnm{{de Koning}}, \binits{C.A.}},
\bauthor{\bsnm{{Nitta}}, \binits{N.}},
\bauthor{\bsnm{{Rollett}}, \binits{T.}},
\bauthor{\bsnm{{Farrugia}}, \binits{C.J.}},
\bauthor{\bsnm{{Forsyth}}, \binits{R.J.}},
\bauthor{\bsnm{{Jackson}}, \binits{B.V.}},
\bauthor{\bsnm{{Jensen}}, \binits{E.A.}},
\bauthor{\bsnm{{Kilpua}}, \binits{E.K.J.}},
\bauthor{\bsnm{{Odstrcil}}, \binits{D.}},
\bauthor{\bsnm{{Webb}}, \binits{D.F.}}:
\byear{2012},
\batitle{{An Analysis of the Origin and Propagation of the Multiple Coronal
  Mass Ejections of 2010 August 1}}.
\bjtitle{\apj}
\bvolume{750},
\bfpage{45}.
\doiurl{10.1088/0004-637X/750/1/45}.
\adsurl{2012ApJ...750...45H}.
\end{barticle}
\endbibitem

\bibitem[\protect\citeauthoryear{{Harrison}
  \textit{et~al.}}{2018}]{2018Harrison}
\begin{barticle}
\bauthor{\bsnm{{Harrison}}, \binits{R.A.}},
\bauthor{\bsnm{{Davies}}, \binits{J.A.}},
\bauthor{\bsnm{{Barnes}}, \binits{D.}},
\bauthor{\bsnm{{Byrne}}, \binits{J.P.}},
\bauthor{\bsnm{{Perry}}, \binits{C.H.}},
\bauthor{\bsnm{{Bothmer}}, \binits{V.}},
\bauthor{\bsnm{{Eastwood}}, \binits{J.P.}},
\bauthor{\bsnm{{Gallagher}}, \binits{P.T.}},
\bauthor{\bsnm{{Kilpua}}, \binits{E.K.J.}},
\bauthor{\bsnm{{M{\"o}stl}}, \binits{C.}},
\bauthor{\bsnm{{Rodriguez}}, \binits{L.}},
\bauthor{\bsnm{{Rouillard}}, \binits{A.P.}},
\bauthor{\bsnm{{Odstr{\v c}il}}, \binits{D.}}:
\byear{2018},
\batitle{{CMEs in the Heliosphere: I. A Statistical Analysis of the
  Observational Properties of CMEs Detected in the Heliosphere from 2007 to
  2017 by STEREO/HI-1}}.
\bjtitle{\solphys}
\bvolume{293},
\bfpage{77}.
\doiurl{10.1007/s11207-018-1297-2}.
\adsurl{2018SoPh..293...77H}.
\end{barticle}
\endbibitem

\bibitem[\protect\citeauthoryear{Hess \textit{et~al.}}{2020}]{2020Hess}
\begin{botherref}
\oauthor{\bsnm{Hess}, \binits{P.}},
\oauthor{\bsnm{Rouillard}, \binits{A.}},
\oauthor{\bsnm{Kouloumvakos}, \binits{A.}},
\oauthor{\bsnm{Liewer}, \binits{P.C.}},
\oauthor{\bsnm{Zhang}, \binits{J.}},
\oauthor{\bsnm{Dhakal}, \binits{S.}},
\oauthor{\bsnm{Stenborg}, \binits{G.}},
\oauthor{\bsnm{Colaninno}, \binits{R.C.}},
\oauthor{\bsnm{Howard}, \binits{R.A.}}:
2020,
Wispr imaging of a pristine cme.
\end{botherref}
\endbibitem

\bibitem[\protect\citeauthoryear{{Howard} \textit{et~al.}}{2008}]{2008Howard}
\begin{barticle}
\bauthor{\bsnm{{Howard}}, \binits{R.A.}},
\bauthor{\bsnm{{Moses}}, \binits{J.D.}},
\bauthor{\bsnm{{Vourlidas}}, \binits{A.}},
\bauthor{\bsnm{{Newmark}}, \binits{J.S.}},
\bauthor{\bsnm{{Socker}}, \binits{D.G.}},
\bauthor{\bsnm{{Plunkett}}, \binits{S.P.}},
\bauthor{\bsnm{{Korendyke}}, \binits{C.M.}},
\bauthor{\bsnm{{Cook}}, \binits{J.W.}},
\bauthor{\bsnm{{Hurley}}, \binits{A.}},
\bauthor{\bsnm{{Davila}}, \binits{J.M.}},
\bauthor{\bsnm{{Thompson}}, \binits{W.T.}},
\bauthor{\bsnm{{St Cyr}}, \binits{O.C.}},
\bauthor{\bsnm{{Mentzell}}, \binits{E.}},
\bauthor{\bsnm{{Mehalick}}, \binits{K.}},
\bauthor{\bsnm{{Lemen}}, \binits{J.R.}},
\bauthor{\bsnm{{Wuelser}}, \binits{J.P.}},
\bauthor{\bsnm{{Duncan}}, \binits{D.W.}},
\bauthor{\bsnm{{Tarbell}}, \binits{T.D.}},
\bauthor{\bsnm{{Wolfson}}, \binits{C.J.}},
\bauthor{\bsnm{{Moore}}, \binits{A.}},
\bauthor{\bsnm{{Harrison}}, \binits{R.A.}},
\bauthor{\bsnm{{Waltham}}, \binits{N.R.}},
\bauthor{\bsnm{{Lang}}, \binits{J.}},
\bauthor{\bsnm{{Davis}}, \binits{C.J.}},
\bauthor{\bsnm{{Eyles}}, \binits{C.J.}},
\bauthor{\bsnm{{Mapson-Menard}}, \binits{H.}},
\bauthor{\bsnm{{Simnett}}, \binits{G.M.}},
\bauthor{\bsnm{{Halain}}, \binits{J.P.}},
\bauthor{\bsnm{{Defise}}, \binits{J.M.}},
\bauthor{\bsnm{{Mazy}}, \binits{E.}},
\bauthor{\bsnm{{Rochus}}, \binits{P.}},
\bauthor{\bsnm{{Mercier}}, \binits{R.}},
\bauthor{\bsnm{{Ravet}}, \binits{M.F.}},
\bauthor{\bsnm{{Delmotte}}, \binits{F.}},
\bauthor{\bsnm{{Auchere}}, \binits{F.}},
\bauthor{\bsnm{{Delaboudiniere}}, \binits{J.P.}},
\bauthor{\bsnm{{Bothmer}}, \binits{V.}},
\bauthor{\bsnm{{Deutsch}}, \binits{W.}},
\bauthor{\bsnm{{Wang}}, \binits{D.}},
\bauthor{\bsnm{{Rich}}, \binits{N.}},
\bauthor{\bsnm{{Cooper}}, \binits{S.}},
\bauthor{\bsnm{{Stephens}}, \binits{V.}},
\bauthor{\bsnm{{Maahs}}, \binits{G.}},
\bauthor{\bsnm{{Baugh}}, \binits{R.}},
\bauthor{\bsnm{{McMullin}}, \binits{D.}},
\bauthor{\bsnm{{Carter}}, \binits{T.}}:
\byear{2008},
\batitle{{Sun Earth Connection Coronal and Heliospheric Investigation
  (SECCHI)}}.
\bjtitle{\ssr}
\bvolume{136},
\bfpage{67}.
\doiurl{10.1007/s11214-008-9341-4}.
\adsurl{2008SSRv..136...67H}.
\end{barticle}
\endbibitem

\bibitem[\protect\citeauthoryear{Howard \textit{et~al.}}{2013}]{2013Howard}
\begin{barticle}
\bauthor{\bsnm{Howard}, \binits{R.}},
\bauthor{\bsnm{Vourlidas}, \binits{A.}},
\bauthor{\bsnm{Korendyke}, \binits{C.}},
\bauthor{\bsnm{Plunkett}, \binits{S.}},
\bauthor{\bsnm{Carter}, \binits{M.}},
\bauthor{\bsnm{Wang}, \binits{D.}},
\bauthor{\bsnm{Rich}, \binits{N.}},
\bauthor{\bsnm{Mcmullin}, \binits{D.}},
\bauthor{\bsnm{Lynch}, \binits{S.}},
\bauthor{\bsnm{Thurn}, \binits{A.}},
\bauthor{\bsnm{Clifford}, \binits{G.}},
\bauthor{\bsnm{Socker}, \binits{D.}},
\bauthor{\bsnm{Thernisien}, \binits{A.}},
\bauthor{\bsnm{Chua}, \binits{D.}},
\bauthor{\bsnm{Linton}, \binits{M.}},
\bauthor{\bsnm{Keller}, \binits{D.}},
\bauthor{\bsnm{Janesick}, \binits{J.}},
\bauthor{\bsnm{Tower}, \binits{J.}},
\bauthor{\bsnm{Grygon}, \binits{M.}},
\bauthor{\bsnm{Lamy}, \binits{P.}}:
\byear{2013},
\batitle{The solar orbiter imager (solohi) instrument for the solar orbiter
  mission}.
\bjtitle{Proc SPIE}
\bvolume{8862}.
\doiurl{10.1117/12.2027657}.
\end{barticle}
\endbibitem

\bibitem[\protect\citeauthoryear{Isavnin, Vourlidas, and
  Kilpua}{2014}]{2014Isavnin}
\begin{barticle}
\bauthor{\bsnm{Isavnin}, \binits{A.}},
\bauthor{\bsnm{Vourlidas}, \binits{A.}},
\bauthor{\bsnm{Kilpua}, \binits{E.K.J.}}:
\byear{2014},
\batitle{Three-dimensional evolution of flux-rope cmes and its relation to the
  local orientation of the heliospheric current sheet}.
\bjtitle{Solar Physics}
\bvolume{289}(\bissue{6}),
\bfpage{2141}.
\doiurl{10.1007/s11207-013-0468-4}.
\end{barticle}
\endbibitem

\bibitem[\protect\citeauthoryear{{Kahler} and {Webb}}{2007}]{2007Kahler}
\begin{barticle}
\bauthor{\bsnm{{Kahler}}, \binits{S.W.}},
\bauthor{\bsnm{{Webb}}, \binits{D.F.}}:
\byear{2007},
\batitle{{V arc interplanetary coronal mass ejections observed with the Solar
  Mass Ejection Imager}}.
\bjtitle{\jgr}
\bvolume{112},
\bfpage{A09103}.
\doiurl{10.1029/2007JA012358}.
\adsurl{2007JGRA..112.9103K}.
\end{barticle}
\endbibitem

\bibitem[\protect\citeauthoryear{{Kilpua} \textit{et~al.}}{2005}]{2005Huttunen}
\begin{barticle}
\bauthor{\bsnm{{Kilpua}}, \binits{E.K.J.}},
\bauthor{\bsnm{{Schwenn}}, \binits{R.}},
\bauthor{\bsnm{{Bothmer}}, \binits{V.}},
\bauthor{\bsnm{{Koskinen}}, \binits{H.E.J.}}:
\byear{2005},
\batitle{{Properties and geoeffectiveness of magnetic clouds in the rising,
  maximum and early declining phases of solar cycle 23}}.
\bjtitle{Annales Geophysicae}
\bvolume{23},
\bfpage{4491}.
\end{barticle}
\endbibitem

\bibitem[\protect\citeauthoryear{Kilpua \textit{et~al.}}{2009}]{2008Kilpua}
\begin{barticle}
\bauthor{\bsnm{Kilpua}, \binits{E.K.J.}},
\bauthor{\bsnm{Liewer}, \binits{P.C.}},
\bauthor{\bsnm{Farrugia}, \binits{C.}},
\bauthor{\bsnm{Luhmann}, \binits{J.G.}},
\bauthor{\bsnm{M{\"o}stl}, \binits{C.}},
\bauthor{\bsnm{Li}, \binits{Y.}},
\bauthor{\bsnm{Liu}, \binits{Y.}},
\bauthor{\bsnm{Lynch}, \binits{B.}},
\bauthor{\bsnm{Russel}, \binits{C.T.}},
\bauthor{\bsnm{Vourlidas}, \binits{A.}},
\bauthor{\bsnm{Acuna}, \binits{M.H.}},
\bauthor{\bsnm{Galvin}, \binits{A.B.}},
\bauthor{\bsnm{Larson}, \binits{D.}},
\bauthor{\bsnm{A.}, \binits{S.J.}}:
\byear{2009},
\batitle{Multispacecraft observations of magnetic clouds and their solar
  origins between 19 and 23 may 2007}.
\bjtitle{Solar Physics}
\bvolume{254}(\bissue{2}).
\doiurl{10.1007/s11207-008-9300-y}.
\end{barticle}
\endbibitem

\bibitem[\protect\citeauthoryear{Kilpua \textit{et~al.}}{2012}]{2012aKilpua}
\begin{barticle}
\bauthor{\bsnm{Kilpua}, \binits{E.K.J.}},
\bauthor{\bsnm{Mierla}, \binits{M.}},
\bauthor{\bsnm{Rodriguez}, \binits{L.}},
\bauthor{\bsnm{Zhukov}, \binits{A.N.}},
\bauthor{\bsnm{Srivastava}, \binits{N.}},
\bauthor{\bsnm{West}, \binits{M.J.}}:
\byear{2012},
\batitle{Estimating travel times of coronal mass ejections to 1 au using
  multi-spacecraft coronagraph data}.
\bjtitle{Solar Physics}
\bvolume{279}(\bissue{2}),
\bfpage{477}.
\doiurl{10.1007/s11207-012-0005-x}.
\end{barticle}
\endbibitem

\bibitem[\protect\citeauthoryear{{Kilpua} \textit{et~al.}}{2017}]{2017bKilpua}
\begin{barticle}
\bauthor{\bsnm{{Kilpua}}, \binits{E.K.J.}},
\bauthor{\bsnm{{Balogh}}, \binits{A.}},
\bauthor{\bsnm{{von Steiger}}, \binits{R.}},
\bauthor{\bsnm{{Liu}}, \binits{Y.D.}}:
\byear{2017},
\batitle{{Geoeffective Properties of Solar Transients and Stream Interaction
  Regions}}.
\bjtitle{\ssr}
\bvolume{212}(\bissue{3-4}),
\bfpage{1271}.
\doiurl{10.1007/s11214-017-0411-3}.
\adsurl{https://ui.adsabs.harvard.edu/abs/2017SSRv..212.1271K}.
\end{barticle}
\endbibitem

\bibitem[\protect\citeauthoryear{Kraft, Puschmann, and
  Luntama}{2017}]{2017Kraft}
\begin{bchapter}
\bauthor{\bsnm{Kraft}, \binits{S.}},
\bauthor{\bsnm{Puschmann}, \binits{K.G.}},
\bauthor{\bsnm{Luntama}, \binits{J.P.}}:
\byear{2017},
\bctitle{{Remote sensing optical instrumentation for enhanced space weather
  monitoring from the L1 and L5 Lagrange points}}.
In: \beditor{\bsnm{Cugny}, \binits{B.}},
\beditor{\bsnm{Karafolas}, \binits{N.}},
\beditor{\bsnm{Sodnik}, \binits{Z.}} (eds.)
\bbtitle{International Conference on Space Optics — ICSO 2016}
\bseriesno{10562},
\bpublisher{SPIE}, \blocation{???},
\bfpage{115 }.
\bcomment{International Society for Optics and Photonics}.
\doiurl{10.1117/12.2296100}.
\burl{https://doi.org/10.1117/12.2296100}.
\end{bchapter}
\endbibitem

\bibitem[\protect\citeauthoryear{{Liu}, {Richardson}, and
  {Belcher}}{2005}]{2005Liu}
\begin{barticle}
\bauthor{\bsnm{{Liu}}, \binits{Y.}},
\bauthor{\bsnm{{Richardson}}, \binits{J.D.}},
\bauthor{\bsnm{{Belcher}}, \binits{J.W.}}:
\byear{2005},
\batitle{{A statistical study of the properties of interplanetary coronal mass
  ejections from 0.3 to 5.4 AU}}.
\bjtitle{\planss}
\bvolume{53},
\bfpage{3}.
\doiurl{10.1016/j.pss.2004.09.023}.
\adsurl{2005P\%26SS...53....3L}.
\end{barticle}
\endbibitem

\bibitem[\protect\citeauthoryear{Liu \textit{et~al.}}{2013}]{2013Liu}
\begin{barticle}
\bauthor{\bsnm{Liu}, \binits{Y.D.}},
\bauthor{\bsnm{Luhmann}, \binits{J.G.}},
\bauthor{\bsnm{Lugaz}, \binits{N.}},
\bauthor{\bsnm{Möstl}, \binits{C.}},
\bauthor{\bsnm{Davies}, \binits{J.A.}},
\bauthor{\bsnm{Bale}, \binits{S.D.}},
\bauthor{\bsnm{Lin}, \binits{R.P.}}:
\byear{2013},
\batitle{{ON} {SUN}-{TO}-{EARTH} {PROPAGATION} {OF} {CORONAL} {MASS}
  {EJECTIONS}}.
\bjtitle{The Astrophysical Journal}
\bvolume{769}(\bissue{1}),
\bfpage{45}.
\doiurl{10.1088/0004-637x/769/1/45}.
\burl{https://doi.org/10.1088/0004-637x/769/1/45}.
\end{barticle}
\endbibitem

\bibitem[\protect\citeauthoryear{{Liu} \textit{et~al.}}{2014}]{2014Liu}
\begin{barticle}
\bauthor{\bsnm{{Liu}}, \binits{Y.D.}},
\bauthor{\bsnm{{Luhmann}}, \binits{J.G.}},
\bauthor{\bsnm{{Kadic}}, \binits{P.}},
\bauthor{\bsnm{{Kilpua}}, \binits{E.K.J.}},
\bauthor{\bsnm{{Lugaz}}, \binits{N.}},
\bauthor{\bsnm{{Nitta}}, \binits{N.V.}},
\bauthor{\bsnm{{M{\"o}stl}}, \binits{C.}},
\bauthor{\bsnm{{Lavraud}}, \binits{B.}},
\bauthor{\bsnm{{Bale}}, \binits{S.D.}},
\bauthor{\bsnm{{Farrugia}}, \binits{C.J.}},
\bauthor{\bsnm{{Galvin}}, \binits{A.B.}}:
\byear{2014},
\batitle{{Observations of an extreme storm in interplanetary space caused by
  successive coronal mass ejections}}.
\bjtitle{Nature Communications}
\bvolume{5}.
\doiurl{10.1038/ncomms4481}.
\end{barticle}
\endbibitem

\bibitem[\protect\citeauthoryear{{Liu} \textit{et~al.}}{2010}]{2010Liu}
\begin{barticle}
\bauthor{\bsnm{{Liu}}, \binits{Y.}},
\bauthor{\bsnm{{Davies}}, \binits{J.A.}},
\bauthor{\bsnm{{Luhmann}}, \binits{J.G.}},
\bauthor{\bsnm{{Vourlidas}}, \binits{A.}},
\bauthor{\bsnm{{Bale}}, \binits{S.D.}},
\bauthor{\bsnm{{Lin}}, \binits{R.P.}}:
\byear{2010},
\batitle{{Geometric Triangulation of Imaging Observations to Track Coronal Mass
  Ejections Continuously Out to 1 AU}}.
\bjtitle{\apjl}
\bvolume{710},
\bfpage{L82}.
\doiurl{10.1088/2041-8205/710/1/L82}.
\adsurl{2010ApJ...710L..82L}.
\end{barticle}
\endbibitem

\bibitem[\protect\citeauthoryear{{Lugaz}}{2010}]{2010bLugaz}
\begin{barticle}
\bauthor{\bsnm{{Lugaz}}, \binits{N.}}:
\byear{2010},
\batitle{{Accuracy and Limitations of Fitting and Stereoscopic Methods to
  Determine the Direction of Coronal Mass Ejections from Heliospheric Imagers
  Observations}}.
\bjtitle{\solphys}
\bvolume{267},
\bfpage{411}.
\doiurl{10.1007/s11207-010-9654-9}.
\adsurl{2010SoPh..267..411L}.
\end{barticle}
\endbibitem

\bibitem[\protect\citeauthoryear{{Lugaz}, {Vourlidas}, and
  {Roussev}}{2009}]{2009Lugaz}
\begin{barticle}
\bauthor{\bsnm{{Lugaz}}, \binits{N.}},
\bauthor{\bsnm{{Vourlidas}}, \binits{A.}},
\bauthor{\bsnm{{Roussev}}, \binits{I.I.}}:
\byear{2009},
\batitle{{Deriving the radial distances of wide coronal mass ejections from
  elongation measurements in the heliosphere - application to CME-CME
  interaction}}.
\bjtitle{\ag}
\bvolume{27},
\bfpage{3479}.
\doiurl{10.5194/angeo-27-3479-2009}.
\adsurl{2009AnGeo..27.3479L}.
\end{barticle}
\endbibitem

\bibitem[\protect\citeauthoryear{Lugaz \textit{et~al.}}{2010}]{2010aLugaz}
\begin{barticle}
\bauthor{\bsnm{Lugaz}, \binits{N.}},
\bauthor{\bsnm{Hernandez-Charpak}, \binits{J.N.}},
\bauthor{\bsnm{Roussev}, \binits{I.I.}},
\bauthor{\bsnm{Davis}, \binits{C.J.}},
\bauthor{\bsnm{Vourlidas}, \binits{A.}},
\bauthor{\bsnm{Davies}, \binits{J.A.}}:
\byear{2010},
\batitle{Determining the azimuthal properties of coronal mass ejections from
  multi-spacecraft remote-sensing observations with stereo secchi}.
\bjtitle{The Astrophysical Journal}
\bvolume{715}(\bissue{1}),
\bfpage{493}.
\end{barticle}
\endbibitem

\bibitem[\protect\citeauthoryear{Lugaz \textit{et~al.}}{2011}]{2011Lugaz}
\begin{barticle}
\bauthor{\bsnm{Lugaz}, \binits{N.}},
\bauthor{\bsnm{Downs}, \binits{C.}},
\bauthor{\bsnm{Shibata}, \binits{K.}},
\bauthor{\bsnm{Roussev}, \binits{I.I.}},
\bauthor{\bsnm{Asai}, \binits{A.}},
\bauthor{\bsnm{Gombosi}, \binits{T.I.}}:
\byear{2011},
\batitle{{Numerical} {Investigation} {of} a {Coronal} {Mass} {Ejection} {from}
  {an} {Anemone} {Active} {Region}: {Reconnection} {and} {Deflection} {of}
  {the} 2005 {August} 22 {Eruption}}.
\bjtitle{The Astrophysical Journal}
\bvolume{738}(\bissue{2}),
\bfpage{127}.
\doiurl{10.1088/0004-637x/738/2/127}.
\burl{https://doi.org/10.1088/0004-637x/738/2/F127}.
\end{barticle}
\endbibitem

\bibitem[\protect\citeauthoryear{Lugaz \textit{et~al.}}{2012}]{2012Lugaz}
\begin{barticle}
\bauthor{\bsnm{Lugaz}, \binits{N.}},
\bauthor{\bsnm{Farrugia}, \binits{C.J.}},
\bauthor{\bsnm{Davies}, \binits{J.A.}},
\bauthor{\bsnm{M\"ostl}, \binits{C.}},
\bauthor{\bsnm{Davis}, \binits{C.J.}},
\bauthor{\bsnm{Roussev}, \binits{I.I.}},
\bauthor{\bsnm{Temmer}, \binits{M.}}:
\byear{2012},
\batitle{The deflection of the two interacting coronal mass ejections of 2010
  may 23-24 as revealed by combined in siyu measurements and heliospheric
  imaging}.
\bjtitle{The Astrophysical Journal}
\bvolume{759}(\bissue{1}),
\bfpage{68}.
\doiurl{10.1088/0004-637x/759/1/68}.
\end{barticle}
\endbibitem

\bibitem[\protect\citeauthoryear{Lugaz \textit{et~al.}}{2017}]{2017Lugaz}
\begin{barticle}
\bauthor{\bsnm{Lugaz}, \binits{N.}},
\bauthor{\bsnm{Temmer}, \binits{M.}},
\bauthor{\bsnm{Wang}, \binits{Y.}},
\bauthor{\bsnm{Farrugia}, \binits{C.J.}}:
\byear{2017},
\batitle{The interaction of successive coronal mass ejections: A review}.
\bjtitle{Solar Physics}
\bvolume{292}(\bissue{4}),
\bfpage{64}.
\doiurl{10.1007/s11207-017-1091-6}.
\burl{https://doi.org/10.1007/s11207-017-1091-6}.
\end{barticle}
\endbibitem

\bibitem[\protect\citeauthoryear{Manchester
  \textit{et~al.}}{2017}]{2017Manchester}
\begin{barticle}
\bauthor{\bsnm{Manchester}, \binits{W.}},
\bauthor{\bsnm{Kilpua}, \binits{E.K.J.}},
\bauthor{\bsnm{Liu}, \binits{Y.D.}},
\bauthor{\bsnm{Lugaz}, \binits{N.}},
\bauthor{\bsnm{Riley}, \binits{P.}},
\bauthor{\bsnm{T{\"o}r{\"o}k}, \binits{T.}},
\bauthor{\bsnm{Vr{\v{s}}nak}, \binits{B.}}:
\byear{2017},
\batitle{The physical processes of cme/icme evolution}.
\bjtitle{Space Science Reviews}
\bvolume{212}(\bissue{3}),
\bfpage{1159}.
\doiurl{10.1007/s11214-017-0394-0}.
\burl{https://doi.org/10.1007/s11214-017-0394-0}.
\end{barticle}
\endbibitem

\bibitem[\protect\citeauthoryear{{M{\"o}stl} and {Davies}}{2013}]{2013Mostl}
\begin{barticle}
\bauthor{\bsnm{{M{\"o}stl}}, \binits{C.}},
\bauthor{\bsnm{{Davies}}, \binits{J.A.}}:
\byear{2013},
\batitle{{Speeds and Arrival Times of Solar Transients Approximated by
  Self-similar Expanding Circular Fronts}}.
\bjtitle{\solphys}
\bvolume{285},
\bfpage{411}.
\doiurl{10.1007/s11207-012-9978-8}.
\adsurl{2013SoPh..285..411M}.
\end{barticle}
\endbibitem

\bibitem[\protect\citeauthoryear{M\"ostl \textit{et~al.}}{2008}]{2008Mostl}
\begin{barticle}
\bauthor{\bsnm{M\"ostl}, \binits{C.}},
\bauthor{\bsnm{Miklenic}, \binits{C.}},
\bauthor{\bsnm{Farrugia}, \binits{C.J.}},
\bauthor{\bsnm{Temmer}, \binits{M.}},
\bauthor{\bsnm{Veronig}, \binits{A.}},
\bauthor{\bsnm{Galvin}, \binits{A.B.}},
\bauthor{\bsnm{Vr\v{s}nak}, \binits{B.}},
\bauthor{\bsnm{Biernat}, \binits{H.K.}}:
\byear{2008},
\batitle{Two-spacecraft reconstruction of a magnetic cloud and comparison to
  its solar source}.
\bjtitle{Annales Geophysicae}
\bvolume{26}(\bissue{10}),
\bfpage{3139}.
\doiurl{10.5194/angeo-26-3139-2008}.
\end{barticle}
\endbibitem

\bibitem[\protect\citeauthoryear{{M{\"o}stl} \textit{et~al.}}{2010}]{2010Mostl}
\begin{barticle}
\bauthor{\bsnm{{M{\"o}stl}}, \binits{C.}},
\bauthor{\bsnm{{Temmer}}, \binits{M.}},
\bauthor{\bsnm{{Rollett}}, \binits{T.}},
\bauthor{\bsnm{{Farrugia}}, \binits{C.J.}},
\bauthor{\bsnm{{Liu}}, \binits{Y.}},
\bauthor{\bsnm{{Veronig}}, \binits{A.M.}},
\bauthor{\bsnm{{Leitner}}, \binits{M.}},
\bauthor{\bsnm{{Galvin}}, \binits{A.B.}},
\bauthor{\bsnm{{Biernat}}, \binits{H.K.}}:
\byear{2010},
\batitle{{STEREO and Wind observations of a fast ICME flank triggering a
  prolonged geomagnetic storm on 5-7 April 2010}}.
\bjtitle{\grl}
\bvolume{37},
\bfpage{L24103}.
\doiurl{10.1029/2010GL045175}.
\adsurl{2010GeoRL..3724103M}.
\end{barticle}
\endbibitem

\bibitem[\protect\citeauthoryear{M\"ostl \textit{et~al.}}{2011}]{2011Mostl}
\begin{barticle}
\bauthor{\bsnm{M\"ostl}, \binits{C.}},
\bauthor{\bsnm{Rollett}, \binits{T.}},
\bauthor{\bsnm{Lugaz}, \binits{N.}},
\bauthor{\bsnm{Farrugia}, \binits{C.J.}},
\bauthor{\bsnm{Davies}, \binits{J.A.}},
\bauthor{\bsnm{Temmer}, \binits{M.}},
\bauthor{\bsnm{Veronig}, \binits{A.M.}},
\bauthor{\bsnm{Harrison}, \binits{R.A.}},
\bauthor{\bsnm{Crothers}, \binits{S.}},
\bauthor{\bsnm{Luhmann}, \binits{J.G.}},
\bauthor{\bsnm{Galvin}, \binits{A.B.}},
\bauthor{\bsnm{Zhang}, \binits{T.L.}},
\bauthor{\bsnm{Baumjohann}, \binits{W.}},
\bauthor{\bsnm{Biernat}, \binits{H.K.}}:
\byear{2011},
\batitle{Arrival time calculation for interplanetary coronal mass ejections
  with circular fronts and application to stereo observations of the 2009
  february 13 eruption}.
\bjtitle{The Astrophysical Journal}
\bvolume{741}(\bissue{1}),
\bfpage{34}.
\doiurl{10.1088/0004-637x/741/1/34}.
\burl{https://doi.org/10.1088/0004-637x/741/1/34}.
\end{barticle}
\endbibitem

\bibitem[\protect\citeauthoryear{{M{\"o}stl} \textit{et~al.}}{2015}]{2015Mostl}
\begin{barticle}
\bauthor{\bsnm{{M{\"o}stl}}, \binits{C.}},
\bauthor{\bsnm{{Rollet}}, \binits{T.}},
\bauthor{\bsnm{{Frahm}}, \binits{R.A.}},
\bauthor{\bsnm{{Liu}}, \binits{Y.D.}},
\bauthor{\bsnm{{Long}}, \binits{D.M.}},
\bauthor{\bsnm{{Colaninno}}, \binits{R.C.}},
\bauthor{\bsnm{A.}, \binits{R.M.}},
\bauthor{\bsnm{M.}, \binits{T.}},
\bauthor{\bsnm{J.}, \binits{F.C.}},
\bauthor{\bsnm{A.}, \binits{P.}},
\bauthor{\bsnm{{Dumbovi{\'c}}}},
\bauthor{\bsnm{M.}, \binits{J.}},
\bauthor{\bsnm{P.}, \binits{D.}},
\bauthor{\bsnm{P.}, \binits{B.}},
\bauthor{\bsnm{A.}, \binits{D.}},
\bauthor{\bsnm{E.}, \binits{K.}},
\bauthor{\bsnm{L.}, \binits{M.M.}},
\bauthor{\bsnm{B.}, \binits{V.}}:
\byear{2015},
\batitle{{Strong coronal channelling and interplanetary evolution of a solar
  storm up to Earth and Mars}}.
\bjtitle{Nature Communications}
\bvolume{6}.
\doiurl{10.1038/ncomms8135}.
\end{barticle}
\endbibitem

\bibitem[\protect\citeauthoryear{{M{\"o}stl} \textit{et~al.}}{2017}]{2017Mostl}
\begin{barticle}
\bauthor{\bsnm{{M{\"o}stl}}, \binits{C.}},
\bauthor{\bsnm{{Isavnin}}, \binits{A.}},
\bauthor{\bsnm{{Boakes}}, \binits{P.D.}},
\bauthor{\bsnm{{Kilpua}}, \binits{E.K.J.}},
\bauthor{\bsnm{{Davies}}, \binits{J.A.}},
\bauthor{\bsnm{{Harrison}}, \binits{R.A.}},
\bauthor{\bsnm{{Barnes}}, \binits{D.}},
\bauthor{\bsnm{{Krupar}}, \binits{V.}},
\bauthor{\bsnm{{Eastwood}}, \binits{J.P.}},
\bauthor{\bsnm{{Good}}, \binits{S.W.}},
\bauthor{\bsnm{{Forsyth}}, \binits{R.J.}},
\bauthor{\bsnm{{Bothmer}}, \binits{V.}},
\bauthor{\bsnm{{Reiss}}, \binits{M.A.}},
\bauthor{\bsnm{{Amerstorfer}}, \binits{T.}},
\bauthor{\bsnm{{Winslow}}, \binits{R.M.}},
\bauthor{\bsnm{{Anderson}}, \binits{B.J.}},
\bauthor{\bsnm{{Philpott}}, \binits{L.C.}},
\bauthor{\bsnm{{Rodriguez}}, \binits{L.}},
\bauthor{\bsnm{{Rouillard}}, \binits{A.P.}},
\bauthor{\bsnm{{Gallagher}}, \binits{P.}},
\bauthor{\bsnm{{Nieves-Chinchilla}}, \binits{T.}},
\bauthor{\bsnm{{Zhang}}, \binits{T.L.}}:
\byear{2017},
\batitle{{Modeling observations of solar coronal mass ejections with
  heliospheric imagers verified with the Heliophysics System Observatory}}.
\bjtitle{Space Weather}
\bvolume{15},
\bfpage{955}.
\doiurl{10.1002/2017SW001614}.
\adsurl{2017SpWea..15..955M}.
\end{barticle}
\endbibitem

\bibitem[\protect\citeauthoryear{M\"ostl \textit{et~al.}}{2018}]{2018Mostl}
\begin{barticle}
\bauthor{\bsnm{M\"ostl}, \binits{C.}},
\bauthor{\bsnm{Amerstorfer}, \binits{T.}},
\bauthor{\bsnm{Palmerio}, \binits{E.}},
\bauthor{\bsnm{Isavnin}, \binits{A.}},
\bauthor{\bsnm{Farrugia}, \binits{C.J.}},
\bauthor{\bsnm{Lowder}, \binits{C.}},
\bauthor{\bsnm{Winslow}, \binits{R.M.}},
\bauthor{\bsnm{Donnerer}, \binits{J.M.}},
\bauthor{\bsnm{Kilpua}, \binits{E.K.J.}},
\bauthor{\bsnm{Boakes}, \binits{P.D.}}:
\byear{2018},
\batitle{Forward modeling of coronal mass ejection flux ropes in the inner
  heliosphere with 3dcore}.
\bjtitle{Space Weather}
\bvolume{16}(\bissue{3}),
\bfpage{216}.
\doiurl{10.1002/2017SW001735}.
\end{barticle}
\endbibitem

\bibitem[\protect\citeauthoryear{{Palmerio}
  \textit{et~al.}}{2019}]{2019Palmerio}
\begin{barticle}
\bauthor{\bsnm{{Palmerio}}, \binits{E.}},
\bauthor{\bsnm{{Scolini}}, \binits{C.}},
\bauthor{\bsnm{{Barnes}}, \binits{D.}},
\bauthor{\bsnm{{Magdelani\'c}}, \binits{J.}},
\bauthor{\bsnm{{West}}, \binits{M.J.}},
\bauthor{\bsnm{{Zhukov}}, \binits{A.N.}},
\bauthor{\bsnm{{Rodriguez}}, \binits{L.}},
\bauthor{\bsnm{{Mierla}}, \binits{M.}},
\bauthor{\bsnm{{Good}}, \binits{S.W.}},
\bauthor{\bsnm{{Morosan}}, \binits{D.E.}},
\bauthor{\bsnm{{Kilpua}}, \binits{E.K.J.}},
\bauthor{\bsnm{{Pomell}}, \binits{J.}},
\bauthor{\bsnm{{Poedts}}, \binits{S.}}:
\byear{2019},
\batitle{{Multipoint Study of Successive Coronal Mass Ejections Driving
  Moderate Disturbances at 1 AU}}.
\bjtitle{ApJ}
\bvolume{878},
\bfpage{37}.
\doiurl{10.3847/1538-4357/ab1850}.
\adsurl{https://doi.org/10.3847/1538-4357/ab1850}.
\end{barticle}
\endbibitem

\bibitem[\protect\citeauthoryear{Richardson and Cane}{2012}]{2012Richardson}
\begin{botherref}
\oauthor{\bsnm{Richardson}, \binits{I.}},
\oauthor{\bsnm{Cane}, \binits{H.}}:
2012,
Near-earth solar wind flows and related geomagnetic activity during more than
  four solar cycles (1963–2011).
\textit{Space Weather and Space Climate}
\textbf{2}.
\end{botherref}
\endbibitem

\bibitem[\protect\citeauthoryear{Robbrecht, Berghmans, and der
  Linden}{2009}]{2009Robbrecht}
\begin{barticle}
\bauthor{\bsnm{Robbrecht}, \binits{E.}},
\bauthor{\bsnm{Berghmans}, \binits{D.}},
\bauthor{\bparticle{der} \bsnm{Linden}, \binits{R.A.M.V.}}:
\byear{2009},
\batitle{Automated lasco cme catalogue for solar cycle 23: are cmes scale
  invariant?}
\bjtitle{The Astrophysical Journal}
\bvolume{691}(\bissue{2}),
\bfpage{1222}.
\doiurl{10.1088/0004-637x/691/2/1222}.
\end{barticle}
\endbibitem

\bibitem[\protect\citeauthoryear{{Rollett} \textit{et~al.}}{2014}]{2014Rollett}
\begin{barticle}
\bauthor{\bsnm{{Rollett}}, \binits{T.}},
\bauthor{\bsnm{{M{\"o}stl}}, \binits{C.}},
\bauthor{\bsnm{{Temmer}}, \binits{M.}},
\bauthor{\bsnm{{Frahm}}, \binits{R.A.}},
\bauthor{\bsnm{{Davies}}, \binits{J.A.}},
\bauthor{\bsnm{{Veronig}}, \binits{A.M.}},
\bauthor{\bsnm{{Vr{\v s}nak}}, \binits{B.}},
\bauthor{\bsnm{{Amerstorfer}}, \binits{U.V.}},
\bauthor{\bsnm{{Farrugia}}, \binits{C.J.}},
\bauthor{\bsnm{{{\v Z}ic}}, \binits{T.}},
\bauthor{\bsnm{{Zhang}}, \binits{T.L.}}:
\byear{2014},
\batitle{{Combined Multipoint Remote and in situ Observations of the Asymmetric
  Evolution of a Fast Solar Coronal Mass Ejection}}.
\bjtitle{\apjl}
\bvolume{790},
\bfpage{L6}.
\doiurl{10.1088/2041-8205/790/1/L6}.
\adsurl{2014ApJ...790L...6R}.
\end{barticle}
\endbibitem

\bibitem[\protect\citeauthoryear{{Rollett} \textit{et~al.}}{2016}]{2016Rollett}
\begin{barticle}
\bauthor{\bsnm{{Rollett}}, \binits{T.}},
\bauthor{\bsnm{{M{\"o}stl}}, \binits{C.}},
\bauthor{\bsnm{{Isavnin}}, \binits{A.}},
\bauthor{\bsnm{{Davies}}, \binits{J.A.}},
\bauthor{\bsnm{{Kubicka}}, \binits{M.}},
\bauthor{\bsnm{{Amerstorfer}}, \binits{U.V.}},
\bauthor{\bsnm{{Harrison}}, \binits{R.A.}}:
\byear{2016},
\batitle{{ElEvoHI: A Novel CME Prediction Tool for Heliospheric Imaging
  Combining an Elliptical Front with Drag-based Model Fitting}}.
\bjtitle{\apj}
\bvolume{824},
\bfpage{131}.
\doiurl{10.3847/0004-637X/824/2/131}.
\adsurl{2016ApJ...824..131R}.
\end{barticle}
\endbibitem

\bibitem[\protect\citeauthoryear{{Rouillard}
  \textit{et~al.}}{2008}]{2008Rouillard}
\begin{barticle}
\bauthor{\bsnm{{Rouillard}}, \binits{A.P.}},
\bauthor{\bsnm{{Davies}}, \binits{J.A.}},
\bauthor{\bsnm{{Forsyth}}, \binits{R.J.}},
\bauthor{\bsnm{{Rees}}, \binits{A.}},
\bauthor{\bsnm{{Davis}}, \binits{C.J.}},
\bauthor{\bsnm{{Harrison}}, \binits{R.A.}},
\bauthor{\bsnm{{Lockwood}}, \binits{M.}},
\bauthor{\bsnm{{Bewsher}}, \binits{D.}},
\bauthor{\bsnm{{Crothers}}, \binits{S.R.}},
\bauthor{\bsnm{{Eyles}}, \binits{C.J.}},
\bauthor{\bsnm{{Hapgood}}, \binits{M.}},
\bauthor{\bsnm{{Perry}}, \binits{C.H.}}:
\byear{2008},
\batitle{{First imaging of corotating interaction regions using the STEREO
  spacecraft}}.
\bjtitle{\grl}
\bvolume{35},
\bfpage{L10110}.
\doiurl{10.1029/2008GL033767}.
\adsurl{2008GeoRL..3510110R}.
\end{barticle}
\endbibitem

\bibitem[\protect\citeauthoryear{{Sachdeva}
  \textit{et~al.}}{2017}]{2017Sachdeva}
\begin{barticle}
\bauthor{\bsnm{{Sachdeva}}, \binits{N.}},
\bauthor{\bsnm{{Subramanian}}, \binits{P.}},
\bauthor{\bsnm{{Vourlidas}}, \binits{A.}},
\bauthor{\bsnm{{Bothmer}}, \binits{V.}}:
\byear{2017},
\batitle{{CME Dynamics Using STEREO and LASCO Observations: The Relative
  Importance of Lorentz Forces and Solar Wind Drag}}.
\bjtitle{\solphys}
\bvolume{292},
\bfpage{118}.
\doiurl{10.1007/s11207-017-1137-9}.
\adsurl{2017SoPh..292..118S}.
\end{barticle}
\endbibitem

\bibitem[\protect\citeauthoryear{{Savani} \textit{et~al.}}{2009}]{2009Savani}
\begin{barticle}
\bauthor{\bsnm{{Savani}}, \binits{N.P.}},
\bauthor{\bsnm{{Rouillard}}, \binits{A.P.}},
\bauthor{\bsnm{{Davies}}, \binits{J.A.}},
\bauthor{\bsnm{{Owens}}, \binits{M.J.}},
\bauthor{\bsnm{{Forsyth}}, \binits{R.J.}},
\bauthor{\bsnm{{Davis}}, \binits{C.J.}},
\bauthor{\bsnm{{Harrison}}, \binits{R.A.}}:
\byear{2009},
\batitle{{The radial width of a Coronal Mass Ejection between 0.1 and 0.4 AU
  estimated from the Heliospheric Imager on STEREO}}.
\bjtitle{\ag}
\bvolume{27},
\bfpage{4349}.
\doiurl{10.5194/angeo-27-4349-2009}.
\adsurl{2009AnGeo..27.4349S}.
\end{barticle}
\endbibitem

\bibitem[\protect\citeauthoryear{Savani \textit{et~al.}}{2010}]{2010Savani}
\begin{barticle}
\bauthor{\bsnm{Savani}, \binits{N.P.}},
\bauthor{\bsnm{Owens}, \binits{M.J.}},
\bauthor{\bsnm{Rouillard}, \binits{A.P.}},
\bauthor{\bsnm{Forsyth}, \binits{R.J.}},
\bauthor{\bsnm{Davies}, \binits{J.A.}}:
\byear{2010},
\batitle{Observational evidence of a coronal mass ejection distortion directly
  attributable to a structured solar wind}.
\bjtitle{The Astrophysical Journal}
\bvolume{714}(\bissue{1}),
\bfpage{L128}.
\doiurl{10.1088/2041-8205/714/1/l128}.
\burl{https://doi.org/10.1088/2041-8205/714/1/l128}.
\end{barticle}
\endbibitem

\bibitem[\protect\citeauthoryear{Savani \textit{et~al.}}{2011}]{2011Savani}
\begin{botherref}
\oauthor{\bsnm{Savani}, \binits{N.P.}},
\oauthor{\bsnm{Owens}, \binits{M.J.}},
\oauthor{\bsnm{Rouillard}, \binits{A.P.}},
\oauthor{\bsnm{Forsyth}, \binits{R.J.}},
\oauthor{\bsnm{Kusano}, \binits{K.}},
\oauthor{\bsnm{Shiota}, \binits{D.}},
\oauthor{\bsnm{Kataoka}, \binits{R.}},
\oauthor{\bsnm{Jian}, \binits{L.}},
\oauthor{\bsnm{Bothmer}, \binits{V.}}:
2011,
Evolution of coronal mass ejection morphology with increasing heliocentric
  distance. ii. in situ observations.
\textbf{732}(2),
117.
\doiurl{10.1088/0004-637x/732/2/117}.
\end{botherref}
\endbibitem

\bibitem[\protect\citeauthoryear{{Savani} \textit{et~al.}}{2012}]{2012Savani}
\begin{barticle}
\bauthor{\bsnm{{Savani}}, \binits{N.P.}},
\bauthor{\bsnm{{Davies}}, \binits{J.A.}},
\bauthor{\bsnm{{Davis}}, \binits{C.J.}},
\bauthor{\bsnm{{Shiota}}, \binits{D.}},
\bauthor{\bsnm{{Rouillard}}, \binits{A.P.}},
\bauthor{\bsnm{{Owens}}, \binits{M.J.}},
\bauthor{\bsnm{{Kusano}}, \binits{K.}},
\bauthor{\bsnm{{Bothmer}}, \binits{V.}},
\bauthor{\bsnm{{Bamford}}, \binits{S.P.}},
\bauthor{\bsnm{{Lintott}}, \binits{C.J.}},
\bauthor{\bsnm{{Smith}}, \binits{A.}}:
\byear{2012},
\batitle{{Observational Tracking of the 2D Structure of Coronal Mass Ejections
  Between the Sun and 1 AU}}.
\bjtitle{\solphys}
\bvolume{279},
\bfpage{517}.
\doiurl{10.1007/s11207-012-0041-6}.
\adsurl{2012SoPh..279..517S}.
\end{barticle}
\endbibitem

\bibitem[\protect\citeauthoryear{{Sheeley} \textit{et~al.}}{2008}]{2008Sheeley}
\begin{barticle}
\bauthor{\bsnm{{Sheeley}}, \binits{N.R.} \bsuffix{Jr.}},
\bauthor{\bsnm{{Herbst}}, \binits{A.D.}},
\bauthor{\bsnm{{Palatchi}}, \binits{C.A.}},
\bauthor{\bsnm{{Wang}}, \binits{Y.-M.}},
\bauthor{\bsnm{{Howard}}, \binits{R.A.}},
\bauthor{\bsnm{{Moses}}, \binits{J.D.}},
\bauthor{\bsnm{{Vourlidas}}, \binits{A.}},
\bauthor{\bsnm{{Newmark}}, \binits{J.S.}},
\bauthor{\bsnm{{Socker}}, \binits{D.G.}},
\bauthor{\bsnm{{Plunkett}}, \binits{S.P.}},
\bauthor{\bsnm{{Korendyke}}, \binits{C.M.}},
\bauthor{\bsnm{{Burlaga}}, \binits{L.F.}},
\bauthor{\bsnm{{Davila}}, \binits{J.M.}},
\bauthor{\bsnm{{Thompson}}, \binits{W.T.}},
\bauthor{\bsnm{{St Cyr}}, \binits{O.C.}},
\bauthor{\bsnm{{Harrison}}, \binits{R.A.}},
\bauthor{\bsnm{{Davis}}, \binits{C.J.}},
\bauthor{\bsnm{{Eyles}}, \binits{C.J.}},
\bauthor{\bsnm{{Halain}}, \binits{J.P.}},
\bauthor{\bsnm{{Wang}}, \binits{D.}},
\bauthor{\bsnm{{Rich}}, \binits{N.B.}},
\bauthor{\bsnm{{Battams}}, \binits{K.}},
\bauthor{\bsnm{{Esfandiari}}, \binits{E.}},
\bauthor{\bsnm{{Stenborg}}, \binits{G.}}:
\byear{2008},
\batitle{{Heliospheric Images of the Solar Wind at Earth}}.
\bjtitle{\apj}
\bvolume{675},
\bfpage{853}.
\doiurl{10.1086/526422}.
\adsurl{2008ApJ...675..853S}.
\end{barticle}
\endbibitem

\bibitem[\protect\citeauthoryear{{Sheeley} \textit{et~al.}}{1999}]{1999Sheeley}
\begin{barticle}
\bauthor{\bsnm{{Sheeley}}, \binits{N.R.}},
\bauthor{\bsnm{{Walters}}, \binits{J.H.}},
\bauthor{\bsnm{{Wang}}, \binits{Y.-M.}},
\bauthor{\bsnm{{Howard}}, \binits{R.A.}}:
\byear{1999},
\batitle{{Continuous tracking of coronal outflows: Two kinds of coronal mass
  ejections}}.
\bjtitle{\jgr}
\bvolume{104},
\bfpage{24739}.
\doiurl{10.1029/1999JA900308}.
\adsurl{1999JGR...10424739S}.
\end{barticle}
\endbibitem

\bibitem[\protect\citeauthoryear{{St.~Cyr} \textit{et~al.}}{1999}]{1999StCyr}
\begin{barticle}
\bauthor{\bsnm{{St.~Cyr}}, \binits{O.C.}},
\bauthor{\bsnm{{Burkepile}}, \binits{J.T.}},
\bauthor{\bsnm{{Hundhausen}}, \binits{A.J.}},
\bauthor{\bsnm{{Lecinski}}, \binits{A.R.}}:
\byear{1999},
\batitle{{A comparison of ground-based and spacecraft observations of coronal
  mass ejections from 1980-1989}}.
\bjtitle{\jgr}
\bvolume{104},
\bfpage{12493}.
\doiurl{10.1029/1999JA900045}.
\adsurl{1999JGR...10412493S}.
\end{barticle}
\endbibitem

\bibitem[\protect\citeauthoryear{{St.~Cyr} \textit{et~al.}}{2000}]{2000StCyr}
\begin{barticle}
\bauthor{\bsnm{{St.~Cyr}}, \binits{O.C.}},
\bauthor{\bsnm{{Plunkett}}, \binits{S.P.}},
\bauthor{\bsnm{{Michels}}, \binits{D.J.}},
\bauthor{\bsnm{{Paswaters}}, \binits{S.E.}},
\bauthor{\bsnm{{Koomen}}, \binits{M.J.}},
\bauthor{\bsnm{{Simnett}}, \binits{G.M.}},
\bauthor{\bsnm{{Thompson}}, \binits{B.J.}},
\bauthor{\bsnm{{Gurman}}, \binits{J.B.}},
\bauthor{\bsnm{{Schwenn}}, \binits{R.}},
\bauthor{\bsnm{{Webb}}, \binits{D.F.}},
\bauthor{\bsnm{{Hildner}}, \binits{E.}},
\bauthor{\bsnm{{Lamy}}, \binits{P.L.}}:
\byear{2000},
\batitle{{Properties of coronal mass ejections: SOHO LASCO observations from
  January 1996 to June 1998}}.
\bjtitle{\jgr}
\bvolume{105},
\bfpage{18169}.
\doiurl{10.1029/1999JA000381}.
\adsurl{2000JGR...10518169S}.
\end{barticle}
\endbibitem

\bibitem[\protect\citeauthoryear{{Tappin} and {Howard}}{2009}]{2009aTappin}
\begin{barticle}
\bauthor{\bsnm{{Tappin}}, \binits{S.J.}},
\bauthor{\bsnm{{Howard}}, \binits{T.A.}}:
\byear{2009},
\batitle{{Interplanetary Coronal Mass Ejections Observed in the Heliosphere: 1.
  Review of Theory}}.
\bjtitle{\ssr}
\bvolume{147},
\bfpage{31}.
\doiurl{10.1007/s11214-009-9542-5}.
\adsurl{2009SSRv..147...31H}.
\end{barticle}
\endbibitem

\bibitem[\protect\citeauthoryear{{Temmer} \textit{et~al.}}{2014}]{2014Temmer}
\begin{barticle}
\bauthor{\bsnm{{Temmer}}, \binits{M.}},
\bauthor{\bsnm{{Veronig}}, \binits{A.M.}},
\bauthor{\bsnm{{Peinhart}}, \binits{V.}},
\bauthor{\bsnm{{Vr{\v s}nak}}, \binits{B.}}:
\byear{2014},
\batitle{{Asymmetry in the CME-CME Interaction Process for the Events from 2011
  February 14-15}}.
\bjtitle{\apj}
\bvolume{785},
\bfpage{85}.
\doiurl{10.1088/0004-637X/785/2/85}.
\adsurl{2014ApJ...785...85T}.
\end{barticle}
\endbibitem

\bibitem[\protect\citeauthoryear{Thernisien, Howard, and
  Vourlidas}{2006}]{2006Thernisien}
\begin{barticle}
\bauthor{\bsnm{Thernisien}, \binits{A.F.R.}},
\bauthor{\bsnm{Howard}, \binits{R.A.}},
\bauthor{\bsnm{Vourlidas}, \binits{A.}}:
\byear{2006},
\batitle{Modeling of flux rope coronal mass ejections}.
\bjtitle{The Astrophysical Journal}
\bvolume{652}(\bissue{1}),
\bfpage{763}.
\doiurl{10.1086/508254}.
\burl{https://doi.org/10.1086/508254}.
\end{barticle}
\endbibitem

\bibitem[\protect\citeauthoryear{{Tousey}, {Howard}, and
  {Koomen}}{1974}]{1974Tousey}
\begin{bchapter}
\bauthor{\bsnm{{Tousey}}, \binits{R.}},
\bauthor{\bsnm{{Howard}}, \binits{R.A.}},
\bauthor{\bsnm{{Koomen}}, \binits{M.J.}}:
\byear{1974},
\bctitle{{The Frequency and Nature of Coronal Transient Events Observed by
  OSO-7*}}.
In: \bbtitle{Bulletin of the American Astronomical Society}
\bseriesno{6},
\bfpage{295}.
\adsurl{1974BAAS....6V.295T}.
\end{bchapter}
\endbibitem

\bibitem[\protect\citeauthoryear{Volpes and Bothmer}{2015}]{2015Volpes}
\begin{barticle}
\bauthor{\bsnm{Volpes}, \binits{L.}},
\bauthor{\bsnm{Bothmer}, \binits{V.}}:
\byear{2015},
\batitle{An application of the stereoscopic self-similar-expansion model to the
  determination of cme-driven shock parameters}.
\bjtitle{Solar Physics}
\bvolume{290}(\bissue{10}),
\bfpage{3005}.
\doiurl{10.1007/s11207-015-0775-z}.
\end{barticle}
\endbibitem

\bibitem[\protect\citeauthoryear{Vourlidas
  \textit{et~al.}}{2011}]{2011Vourlidas}
\begin{barticle}
\bauthor{\bsnm{Vourlidas}, \binits{A.}},
\bauthor{\bsnm{Colaninno}, \binits{R.}},
\bauthor{\bsnm{Nieves-Chinchilla}, \binits{T.}},
\bauthor{\bsnm{Stenborg}, \binits{G.}}:
\byear{2011},
\batitle{The first observation of a rapidly rotating coronal mass ejection in
  the middle corona}.
\bjtitle{The Astrophysical Journal}
\bvolume{733}(\bissue{2}),
\bfpage{L23}.
\doiurl{10.1088/2041-8205/733/2/l23}.
\end{barticle}
\endbibitem

\bibitem[\protect\citeauthoryear{{Vourlidas}
  \textit{et~al.}}{2016}]{2016Vourlidas}
\begin{barticle}
\bauthor{\bsnm{{Vourlidas}}, \binits{A.}},
\bauthor{\bsnm{{Howard}}, \binits{R.A.}},
\bauthor{\bsnm{{Plunkett}}, \binits{S.P.}},
\bauthor{\bsnm{{Korendyke}}, \binits{C.M.}},
\bauthor{\bsnm{{Thernisien}}, \binits{A.F.R.}},
\bauthor{\bsnm{{Wang}}, \binits{D.}},
\bauthor{\bsnm{{Rich}}, \binits{N.}},
\bauthor{\bsnm{{Carter}}, \binits{M.T.}},
\bauthor{\bsnm{{Chua}}, \binits{D.H.}},
\bauthor{\bsnm{{Socker}}, \binits{D.G.}},
\bauthor{\bsnm{{Linton}}, \binits{M.G.}},
\bauthor{\bsnm{{Morrill}}, \binits{J.S.}},
\bauthor{\bsnm{{Lynch}}, \binits{S.}},
\bauthor{\bsnm{{Thurn}}, \binits{A.}},
\bauthor{\bsnm{{Van Duyne}}, \binits{P.}},
\bauthor{\bsnm{{Hagood}}, \binits{R.}},
\bauthor{\bsnm{{Clifford}}, \binits{G.}},
\bauthor{\bsnm{{Grey}}, \binits{P.J.}},
\bauthor{\bsnm{{Velli}}, \binits{M.}},
\bauthor{\bsnm{{Liewer}}, \binits{P.C.}},
\bauthor{\bsnm{{Hall}}, \binits{J.R.}},
\bauthor{\bsnm{{DeJong}}, \binits{E.M.}},
\bauthor{\bsnm{{Mikic}}, \binits{Z.}},
\bauthor{\bsnm{{Rochus}}, \binits{P.}},
\bauthor{\bsnm{{Mazy}}, \binits{E.}},
\bauthor{\bsnm{{Bothmer}}, \binits{V.}},
\bauthor{\bsnm{{Rodmann}}, \binits{J.}}:
\byear{2016},
\batitle{{The Wide-Field Imager for Solar Probe Plus (WISPR)}}.
\bjtitle{\ssr}
\bvolume{204},
\bfpage{83}.
\doiurl{10.1007/s11214-014-0114-y}.
\adsurl{2016SSRv..204...83V}.
\end{barticle}
\endbibitem

\bibitem[\protect\citeauthoryear{Vourlidas
  \textit{et~al.}}{2017}]{2017Vourlidas}
\begin{barticle}
\bauthor{\bsnm{Vourlidas}, \binits{A.}},
\bauthor{\bsnm{Balmaceda}, \binits{L.A.}},
\bauthor{\bsnm{Stenborg}, \binits{G.}},
\bauthor{\bsnm{Lago}, \binits{A.D.}}:
\byear{2017},
\batitle{Multi-viewpoint coronal mass ejection catalog based on stereo cor2
  observations}.
\bjtitle{\apj}
\bvolume{838}(\bissue{2}),
\bfpage{141}.
\end{barticle}
\endbibitem

\bibitem[\protect\citeauthoryear{{Wang} \textit{et~al.}}{2004}]{2004Wang}
\begin{barticle}
\bauthor{\bsnm{{Wang}}, \binits{Y.}},
\bauthor{\bsnm{{Shen}}, \binits{C.}},
\bauthor{\bsnm{{Wang}}, \binits{S.}},
\bauthor{\bsnm{{Ye}}, \binits{P.}}:
\byear{2004},
\batitle{{Deflection of coronal mass ejection in the interplanetary medium}}.
\bjtitle{\solphys}
\bvolume{222}(\bissue{2}).
\doiurl{10.1023/B:SOLA.0000043576.21942.aa}.
\end{barticle}
\endbibitem

\bibitem[\protect\citeauthoryear{Wang \textit{et~al.}}{2014}]{2014Wang}
\begin{barticle}
\bauthor{\bsnm{Wang}, \binits{Y.}},
\bauthor{\bsnm{Wang}, \binits{B.}},
\bauthor{\bsnm{Shen}, \binits{C.}},
\bauthor{\bsnm{Shen}, \binits{F.}},
\bauthor{\bsnm{Lugaz}, \binits{N.}}:
\byear{2014},
\batitle{Deflected propagation of a coronal mass ejection from the corona to
  interplanetary space}.
\bjtitle{Journal of Geophysical Research: Space Physics}
\bvolume{119}(\bissue{7}),
\bfpage{5117}.
\doiurl{10.1002/2013JA019537}.
\end{barticle}
\endbibitem

\bibitem[\protect\citeauthoryear{Webb and Howard}{2012}]{2012Webb}
\begin{barticle}
\bauthor{\bsnm{Webb}, \binits{D.F.}},
\bauthor{\bsnm{Howard}, \binits{T.A.}}:
\byear{2012},
\batitle{Coronal mass ejections: Observations}.
\bjtitle{Living Reviews in Solar Physics}
\bvolume{9}(\bissue{3}).
\doiurl{10.12942/lrsp-2012-3}.
\end{barticle}
\endbibitem

\bibitem[\protect\citeauthoryear{{Yashiro} \textit{et~al.}}{2004}]{2004Yashiro}
\begin{barticle}
\bauthor{\bsnm{{Yashiro}}, \binits{S.}},
\bauthor{\bsnm{{Gopalswamy}}, \binits{N.}},
\bauthor{\bsnm{{Michalek}}, \binits{G.}},
\bauthor{\bsnm{{St.~Cyr}}, \binits{O.C.}},
\bauthor{\bsnm{{Plunkett}}, \binits{S.P.}},
\bauthor{\bsnm{{Rich}}, \binits{N.B.}},
\bauthor{\bsnm{{Howard}}, \binits{R.A.}}:
\byear{2004},
\batitle{{A catalog of white light coronal mass ejections observed by the SOHO
  spacecraft}}.
\bjtitle{Journal of Geophysical Research (Space Physics)}
\bvolume{109},
\bfpage{A07105}.
\doiurl{10.1029/2003JA010282}.
\adsurl{2004JGRA..109.7105Y}.
\end{barticle}
\endbibitem

\bibitem[\protect\citeauthoryear{{Zhang} \textit{et~al.}}{2001}]{2001Zhang}
\begin{barticle}
\bauthor{\bsnm{{Zhang}}, \binits{J.}},
\bauthor{\bsnm{{Dere}}, \binits{K.P.}},
\bauthor{\bsnm{{Howard}}, \binits{R.A.}},
\bauthor{\bsnm{{Kundu}}, \binits{M.R.}},
\bauthor{\bsnm{{White}}, \binits{S.M.}}:
\byear{2001},
\batitle{{On the Temporal Relationship between Coronal Mass Ejections and
  Flares}}.
\bjtitle{\apj}
\bvolume{559},
\bfpage{452}.
\doiurl{10.1086/322405}.
\adsurl{2001ApJ...559..452Z}.
\end{barticle}
\endbibitem

\bibitem[\protect\citeauthoryear{{Zhang} \textit{et~al.}}{2004}]{2004Zhang}
\begin{barticle}
\bauthor{\bsnm{{Zhang}}, \binits{J.}},
\bauthor{\bsnm{{Dere}}, \binits{K.P.}},
\bauthor{\bsnm{{Howard}}, \binits{R.A.}},
\bauthor{\bsnm{{Vourlidas}}, \binits{A.}}:
\byear{2004},
\batitle{{A Study of the Kinematic Evolution of Coronal Mass Ejections}}.
\bjtitle{\apj}
\bvolume{604},
\bfpage{420}.
\doiurl{10.1086/381725}.
\adsurl{2004ApJ...604..420Z}.
\end{barticle}
\endbibitem

\bibitem[\protect\citeauthoryear{Zhang \textit{et~al.}}{2007}]{2007Zhang}
\begin{barticle}
\bauthor{\bsnm{Zhang}, \binits{J.}},
\bauthor{\bsnm{Richardson}, \binits{I.G.}},
\bauthor{\bsnm{Webb}, \binits{D.F.}},
\bauthor{\bsnm{Gopalswamy}, \binits{N.}},
\bauthor{\bsnm{Huttunen}, \binits{E.}},
\bauthor{\bsnm{Kasper}, \binits{J.C.}},
\bauthor{\bsnm{Nitta}, \binits{N.V.}},
\bauthor{\bsnm{Poomvises}, \binits{W.}},
\bauthor{\bsnm{Thompson}, \binits{B.J.}},
\bauthor{\bsnm{Wu}, \binits{C.-C.}},
\bauthor{\bsnm{Yashiro}, \binits{S.}},
\bauthor{\bsnm{Zhukov}, \binits{A.N.}}:
\byear{2007},
\batitle{Solar and interplanetary sources of major geomagnetic storms (dst
  $\leq$ -100 nt) during 1996--2005}.
\bjtitle{\jgr}
\bvolume{112}(\bissue{A10}).
\doiurl{10.1029/2007JA012321}.
\end{barticle}
\endbibitem

\end{thebibliography}

\end{article} 
\end{document}